%% file: 3201binaries.tex
\documentclass{aa}
\usepackage{graphicx}
\usepackage[varg]{txfonts}
\usepackage[colorlinks=false]{hyperref}

\usepackage[load-configurations=abbreviations]{siunitx}
\usepackage{silence}

\DeclareSIUnit\year{yr}
\DeclareSIUnit\astronomicalunit{AU}
\DeclareSIUnit\au{AU}
\DeclareSIUnit\parsec{pc}
\DeclareSIUnit\solarmass{\ensuremath{\textrm{M}_\odot}}
\DeclareSIUnit\solarradius{\ensuremath{\textrm{R}_\odot}}

\newcommand{\gc}{globular cluster}
\newcommand{\Gc}{Globular cluster}
\newcommand{\ngc}[1]{NGC~#1}
\newcommand{\p}{\; .}
\newcommand{\bs}{blue straggler}
\newcommand{\bss}{blue straggler star}

\newcommand{\Bss}{Blue straggler star}
\newcommand{\RRL}{RR Lyrae}
\newcommand{\SXP}{SX Phoenicis}
\newcommand{\Ssg}{Sub-subgiant}
\newcommand{\ssg}{sub-subgiant}
\newcommand{\pampelmuse}{\textsc{PampelMuse}}
\newcommand{\spexxy}{\textsc{Spexxy}}
\newcommand{\joker}{\textsc{The Joker}}
\newcommand{\mocca}{\textsc{MOCCA}}

\newcommand{\ms}{main-sequence}
\newcommand{\changed}[1]{\textbf{#1}}

\newcommand{\obf}{\SI{20.08(22)}{\percent}}
\newcommand{\cbf}{\SI{6.75(72)}{\percent}}
\newcommand{\bbf}{\SI{57.5(79)}{\percent}}
\newcommand{\bh}{\SI{4.53(21)}{\solarmass}}

\newcommand\nfh{\mbox{$^{\mathrm h}$}}% 
\newcommand\nfm{\mbox{$^{\mathrm m}$}}% 

\newcommand{\Autoref}[1]{%
 \begingroup%
 \renewcommand\sectionautorefname{Section}%
 \renewcommand\figureautorefname{Figure}%
 \renewcommand\equationautorefname{Equation}%
 \autoref{#1}%
 \endgroup%
}

\makeatletter
\renewcommand*\aa@pageof{, page \thepage{} of \pageref*{LastPage}}
\makeatother

\begin{document}

\def\sectionautorefname{Sect.}
\def\figureautorefname{Fig.}
\def\equationautorefname{Eq.}
\def\subsectionautorefname{Sect.}

\title{A stellar census in globular clusters with MUSE: Binaries in \ngc{3201}}
%\subtitle{Binary fraction, Blue Stragglers, Sub-Subgiants, \SXP, Black Holes}

\author{Benjamin~Giesers\inst{1}
\and Sebastian~Kamann\inst{2}
\and Stefan~Dreizler\inst{1}
\and Tim-Oliver~Husser\inst{1}
\and Abbas~Askar\inst{3}
\and Fabian~Göttgens\inst{1}
\and Jarle~Brinchmann\inst{4,5}
\and Marilyn~Latour\inst{1}
\and Peter~M.~Weilbacher\inst{6}
\and Martin~Wendt\inst{7}
\and Martin~M.~Roth\inst{6}}

\institute{Institut f\"ur Astrophysik, Georg-August-Universit\"at G\"ottingen, Friedrich-Hund-Platz 1, 37077 G\"ottingen, Germany\\
\email{giesers@astro.physik.uni-goettingen.de}
\and Astrophysics Research Institute, Liverpool John Moores University, 146 Brownlow Hill, Liverpool L3 5RF, United Kingdom
\and Lund Observatory, Department of Astronomy and Theoretical Physics, Lund University, Box 43, SE-221 00 Lund, Sweden
\and Instituto de Astrof{\'\i}sica e Ci{\^e}ncias do Espaço, Universidade do Porto, CAUP, Rua das Estrelas, PT4150-762 Porto, Portugal
\and Leiden Observatory, Leiden University, P.O. Box 9513, 2300 RA, Leiden, The Netherlands
\and Leibniz-Institut f\"ur Astrophysik Potsdam (AIP), An der Sternwarte 16, 14482 Potsdam, Germany
\and Institut f\"ur Physik und Astronomie, Universit\"at
Potsdam, Karl-Liebknecht-Str. 24/25, 14476 Golm, Germany}

\date{Received 2019-06-28; accepted XXXX}

\abstract{We utilize multi-epoch MUSE spectroscopy to study binary stars in the core of the Galactic \gc{} \ngc{3201}. Our sample consists of \num{3553} stars with \num{54883} spectra in total comprising \num{3200} \ms{} stars up to 4 magnitudes below the turn-off. Each star in our sample has between 3 and 63 (with a median of 14) reliable radial velocity measurements within five years of observations. We introduce a statistical method to determine the probability of a star showing radial velocity variations based on the whole inhomogeneous radial velocity sample. Using HST photometry and an advanced dynamical \mocca{} simulation of this specific cluster we overcome observational biases that previous spectroscopic studies had to deal with. This allows us to infer a binary frequency in the MUSE FoV and enables us to deduce the underlying true binary frequency of \cbf{} in \ngc{3201}. The comparison of the MUSE observations with the \mocca{} simulation suggests a significant fraction of primordial binaries. We can also confirm a radial increase of the binary fraction towards the cluster centre due to mass segregation. We discovered that in the core of \ngc{3201} at least \bbf{} of \bss{}s are in a binary system.
For the first time in a study of \gc{}s, we were able to fit Keplerian orbits to a significant sample of 95 binaries. We present the binary system properties of eleven \bss{}s and the connection to \SXP{}-type stars. We show evidence that two \bs{} formation scenarios, the mass transfer in binary (or triple) star systems and the coalescence due to binary-binary interactions, are present in our data. We also describe the binary and spectroscopic properties of four \ssg{} (or red straggler) stars. Furthermore, we discovered two new black hole candidates with minimum masses ($M \sin{i}$) of \SI{7.68(50)}{\solarmass}, \SI{4.4(28)}{\solarmass}, and refine the minimum mass estimate on the already published black hole to \bh{}. These black holes are consistent with an extensive black hole subsystem hosted by \ngc{3201}.
}


\titlerunning{Binaries in \ngc{3201}}

\maketitle

%%%%%%%%%%%%%%%%%%%%%%%%%%%%%%%
\input{3201binaries_content}
%%%%%%%%%%%%%%%%%%%%%%%%%%%%%%%

\bibpunct{(}{)}{;}{a}{}{,} % to follow the A&A style

% for the bibliography, at the end
\bibliographystyle{aa} % style aa.bst
\bibliography{3201binaries.bib} % your references Yourfile.bib

% problems with longtable and \begin{appendix}
\begin{appendix}
\section{Additional tables and figures}
\input{3201binaries_appendix}
\end{appendix}

\end{document}

%% file: 3201binaries_content.tex
\section{Introduction}
\label{sec:introduction}
Multi-star systems are not only responsible for an abundance of extraordinary objects like cataclysmic variables, millisecond pulsars, and low-mass X-ray binaries, but can also be regarded as a source of energy: Embedded in the environment of other stars, like in star clusters, the gravitational energy stored in a multi-star system can be transferred to other stars. In \gc{}s with stellar encounters on short time-scales this energy deposit is for example delaying the core collapse \citep{1989Natur.339...40G}. From a thermodynamic point of view, multi-star systems in \gc{}s act as a heat source. The prerequisite for this mechanism is, of course, that \gc{}s do contain multi-star systems at some point during their lifetime. For the \ms{} field stars $> \SI{50}{\percent}$ \citep{duquennoy1991,duchene2013} in our Milky Way are in multi-star systems. In \gc{}s, binary fractions are typically lower than in the field \citep{milone2012}, but can reach values around $\sim50\%$ \citep[e.g][]{sollima2007,milone2012} in central regions. This radial gradient can be explained by mass segregation and indeed is reproduced in simulations \citep[e.g.][]{2009ApJ...707.1533F,2007ApJ...665..707H}.
The initial fraction of multi-star systems in \gc{}s is changed due to dynamical formation and destruction processes caused by stellar encounters as well as stellar evolution during the (typical long) life time of Galactic \gc{}s.
The primordial fraction of multi-star systems to single stars for \gc{}s therefore cannot be easily inferred from observations \citep{hut1992}. As a result, the initial fraction is poorly constrained and simulations use values in a wide range from $5\%$ \citep{2007ApJ...665..707H} to $100\%$ \citep{2005MNRAS.358..572I}. Current observations are still consistent with this wide range in primordial binary fractions, depending on which orbit distribution is assumed \citep{leigh2015}. %In 1979 there was still no evidence for binary systems in globular clusters \citep{gunn1979}. Due to observational challenges and the nature of predominantly hard binaries, it needed new observational techniques to discover them.

Until today there is only one technique with a high discovery efficiency for multi-star systems, namely high precision photometry \citep{sollima2007, milone2012}. It uses the fact that binary stars with both components contributing to the total brightness of one unresolved source have a position in the \textit{Colour-Magnitude-Diagram} (CMD) differing from single stars. For example, a large number of binaries is located to the brighter (redder) side of the \textit{\ms{}} (MS). This method is sensitive to binaries on the MS with arbitrary orbital period and inclination. However, no constraints on orbital parameters are possible. The advantage is that this method is efficient in terms of observing time and produces large statistical robust samples. The disadvantage is that in terms of individual binary system properties it only has access to the biased mass-ratio parameter. \citet{milone2012} investigated 59 Galactic \gc{}s with this method using the \textit{Hubble Space Telescope} (HST). They studied MS binaries with a mass ratio $q > 0.5$ and found in nearly all \gc{}s a significantly smaller binary frequency than \SI{50}{\percent}. The results show a higher concentration of binaries in the core with a general decreasing binary fraction from the centre to about two core radii by a factor of $\sim 2$. A significant anti-correlation between the cluster binary frequency and its absolute luminosity (mass) was found. Additionally, the authors confirm a significant correlation between the fraction of binaries and the fraction of \textit{\bss{}s} (BSS), indicating a relation between the BSS formation mechanism and binaries.

In the literature there are almost no spectroscopic studies of the binarity in \gc{}s. One systematic search was done on the \gc{} M4 based on \num{5973} spectra of \num{2469} stars by \citet{2009A&A...493..947S}. They discovered 57 binary star candidates and derived a lower limit of the total binary fraction of \SI{3.0(3)}{\percent} for M4.

In light of the important role of binaries in \gc{}s, we designed the observing strategy of the MUSE \gc{} survey \citep{kamann2018} such that we can infer the binary fraction of every cluster in our sample. For all clusters we aim for at least three epochs of every pointing. For a selection of clusters, namely \ngc{104} (47 Tuc), \ngc{3201}, and \ngc{5139} ($\omega$~Cen), we are also able to study individual binary systems in more detail. In this paper we concentrate on the \gc{} \ngc{3201}. \ngc{3201} has, apart \ngc{5139}, the largest half-light and core radius of all \gc{}s in our survey. Its binary fraction within the core radius appears to be relatively high \citep[\SI{12.8(8)}{\percent},][]{milone2012}. In
\citet{giesers}, we reported a quiescent detached stellar-mass black hole with \SI{4.36(41)}{\solarmass} in this cluster, the first black hole found by a blind spectroscopic survey. Based upon our discovery, \ngc{3201} was predicted to harbour a significant population of black holes \citep[see][]{kremer2018,askar2018b}. In our MUSE data, \ngc{3201} has the deepest observations, the most epochs to date, and the least crowding. It is therefore the ideal case for our purpose.

This paper is structured as follows. In \autoref{sec:peculiar} we describe peculiar objects which are important in the context of binary star studies in \ngc{3201}. In \autoref{sec:observations} the observations, data reduction and selection of our final sample are discussed. In \autoref{sec:simulation} we present a \mocca{} simulation of the \gc{} \ngc{3201} that we used to create a mock observation for comparisons and verifying the following method. We introduce a statistical method to determine the variability of radial velocity measurements and use the method to determine the binary fraction of \ngc{3201} in \autoref{sec:stats}. In \autoref{sec:orbits} we explain our usage of the tool \joker{} to find orbital solutions to individual binary systems and present their orbital parameter distributions. We discuss peculiar objects in our data, like \bss{}s, \ssg{}s and black hole candidates in \autoref{sec:discussion}. We end with the conclusions and outlook in \autoref{sec:conclusions}.

\section{Peculiar objects in \gc{}s}
\label{sec:peculiar}
\subsection{\Bss{}s}
In most \gc{}s a collection of \bss{}s can be easily found in the CMD along an extension of the \ms{} beyond the turn-off point. Their positions indicate more massive and younger stars than those of the \ms{}. %As it is believed that almost all stars in \gc{}s are formed at the same time, \bs{}s are a curiosity. Since initial binary stars can influence significantly the evolution of a \gc{} \citep[see][]{goodman1989}, a study of \bs{}s in binary systems could help to get a better understanding on the link between the stellar and dynamical evolution of \gc{}s \citep[see][]{bailyn1995}.
%How is a \bs{} created?
Without continuing star formation due to lacking gas and dust in \gc{}s, mass transfer from a companion is the usual explanation for the formation of a \bs{}.
This mass transfer could be on long time-scales, like in compact binary systems, or on short time-scales, as for collisions of two stars (in a binary system or of a priori not associated stars). However, it is still unclear which of these processes contribute to the formation of \bs{}s and how often. %Additionally it is not clear on which \gc{} structural parameters they depend most \citep{milone2012}.
% But there are some known correlations. \citet{pecci1992} found the ratio $N_\textrm{BSS}/N^\textrm{HB}_\textrm{BSP} \sim 6$ of detected \bs{}s $N_\textrm{BSS}$ and their progeny on the Horizontal Branch $N^\textrm{HB}_\textrm{BSP}$ in \gc{}s. 
\citet{ferraro1997} discovered a bimodal radial distribution of \bs{}s in M3 and suggested that the \bs{}s in the inner cluster are formed by stellar collisions and those in the outer cluster from merging primordial binaries. \citet{sollima2008} identified that the strongest correlation in low-density \gc{}s is between the number of \bs{}s and the binary frequency of the cluster. They suggested that the primordial binary fraction is one of the most important factors for producing \bs{}s and that the collisional channel to form \bs{}s has a very small efficiency in low-density \gc{}s. %\citet{knigge2009} pointed out that there is a strong correlation between the observed number of \bs{}s found in the core of a given \gc{} versus the estimated core mass and the number of core binaries. However they also stated that the total number of \bs{}s does not correlate with the predicted collision rate in a \gc{} which depends directly on the cluster density. They conclude that mass transfer in binary systems seems to be more efficient to form \bs{}s. \citet{ferraro2009} found two distinct parallel \bs{} sequences in the CMD of the core collapsed \gc{} M30. They suggested that the bluer population originates from direct stellar collisions whereas the redder population originates from the evolution of close binaries. \citet{xin2015} confirmed the origin of the redder population from binary evolution. \citet{leigh2013} investigated if the binary frequency contributes to the \bs{} number vs core mass correlation. They found that the core mass alone is a better predictor of \bs{} numbers.

\citet{hypki2016} performed numerical simulations of \gc{}s with various initial conditions. They found that the number of evolutionary \bs{}s (formed in a compact binary system) is not affected by the density of the \gc{}. In contrast, the number of dynamically created \bs{}s (by collisions between stars) correlates with the density. The efficiencies of both channels strongly depend on the initial semi-major axes distributions. Wider initial orbits lead to more dynamically created \bs{}s. Finally, they found in all \gc{} models at the end of the simulation a constant ratio between the number of \bs{}s in binaries and as single stars $R_\textrm{B/S} \sim 0.4$. %Hence, the binary \bs{} formation is dominated by the binary system evolution and the single \bs{} formation is dominated by dynamical interactions.

%Unfortunately we still have an observational discriminability problem of the \bs{} formation channels. Indeed \citet{ferraro2006} found a \bs{} sub-population with a significant depletion of Carbon (C) and Oxygen (O) in comparison to the remaining \bs{} population. But the interpretation in respect to the formation channels is not clear \citep[see][]{ferraro2006,fossati2010}.

A mass distribution of \bs{}s has been estimated by \citet{fiorentino2014} from pulsation properties of \bs{}s in the \gc{} \ngc{6541}. They found a mass of \SI{1.06(9)}{\solarmass}, which is significantly in excess of the cluster \ms{} turn-off mass (\SI{0.75}{\solarmass}). \citet{baldwin2016} found an average mass of \bss{}s of \SI{1.22(12)}{\solarmass} in 19 \gc{}s. \citet{geller2011} found a surprisingly narrow \bs{} companion mass distribution with a mean of \SI{0.53}{\solarmass} in long-period binaries in the open cluster \ngc{188}. They suggested that this conclusively rules out a collisional origin in this cluster, as the collision hypothesis predicts significantly higher companion masses in their model.

\subsection{\Ssg{} stars}
A star in a cluster is called a \ssg{} (SSG) if its position in the CMD is redder than MS turn-off stars (and its binaries) and fainter than normal subgiant stars. These stars are related to the so called red stragglers (RS) which are redder than normal (red) giant stars. For our study this distinction is not important and we call them hereafter \ssg{}s. The evolution of these stars cannot be explained by single star evolution. \citet{geller2017a} describe in detail the demographics of SSGs in open and \gc{}s and conclude that half of their SSG sample are radial velocity binary stars with typical periods less than \SI{15}{\day}, \SI{58}{\percent} are X-ray sources, and \SI{33}{\percent} are H$\alpha$ emitters. The formation scenarios explained in \citet{leiner2017} for SSG and RS are isolated binary evolution, the rapid stripping of a subgiant’s envelope, or stellar collisions. So far, from observations and simulations, no channel can be excluded, but it seems that the binary channel is dominant in \gc{}s \citep{geller2017b}.

\subsection{\SXP{}-type stars}
\SXP{}-type (SXP) stars are low-luminosity pulsators in the classical instability strip with short periods. In \gc{}s they appear as \bss{}s with an enhanced helium content \citep{cohen2012}. SXP stars show a strong period-luminosity relation, from which distances can be measured. The period range is from \SI{0.02}{\day} to \SI{0.4}{\day} with typical periods around $\sim \SI{0.04}{\day} \approx \SI{1}{\hour}$. SXP stars are thought to be formed in binary evolution, but it is not clear whether they are indicative of a particular blue straggler formation channel. The prototype star SX Phoenicis shows a radial velocity amplitude of \SI{19}{\kilo\meter\per\second} due to radial pulsation modes \citep{1993AJ....106.2493K}.

\subsection{Black holes}
Observers are searching for two kinds of black holes in \gc{}s.
On the one hand, we have no conclusive evidence for central intermediate-mass black holes \citep[IMBH, see][]{baumgardt2017,tremou2018}. One of the challenges in searching for IMBHs is that typical signatures, such as a central cusp in the surface brightness profile or the velocity dispersion profile can also be explained by a binary star or stellar-mass black hole population \citep[for Omega Cen, see][]{zocchi2019}. %One candidate indirectly determined by millisecond pulsars is in the \gc{} 47 Tuc \citep{kiziltan2017}.
On the other hand, there are known stellar-mass black holes in Galactic \gc{}s \citep[e.g.][]{strader2012,giesers}.
If they are retained, stellar-mass black holes should quickly migrate to the cluster cores due to mass segregation, and strongly affect the evolution of their host \gc{}s \citep[e.g.][]{kremer2018,sedda2018,askar2018b,askar2019}. However, the retention fraction of black holes in \gc{}s is largely unknown, which also limits our abilities to interprete gravitational wave detections. For \ngc{3201}, \citet{askar2018b} and \citet{kremer2018} predicted that between 100 and more than 200 stellar-mass black holes should still be present in the cluster, with up to ten in a binary system with a \ms{} star.

\section{Observations and data reduction}
\label{sec:observations}

\begin{table}
 \caption{Exposures, observations and integration time per pointing.}
 \label{table:observations}
 \centering
 \begin{tabular}{c c c c}
 \hline\hline
     Pointing & Exposures & Observations & Total time [min.] \\
  \hline
  01 & 49 & 17 & 163 \\
  02 & 49 & 15 & 143 \\
  03 & 44 & 15 & 143 \\
  04 & 36 & 12 & 114 \\
  Deep & 16 & 4 & 160 \\ \hline
  $\sum$ & 194 & 63 & 723 \\
  \hline
 \end{tabular}
\end{table}

The observational challenge in \gc{}s is the crowded field resulting in multiple star blends. For photometric measurements of dense \gc{}s, instruments like the ones attached to the HST have just sufficient spatial resolution to get independent measurements of stars down the \ms{}. In the past, spectroscopic measurements of individual \gc{} stars (with a reasonable spectroscopic resolution) were limited by crowding in the cluster centre. Since 2014 we are observing 27 Galactic \gc{}s (PI: S. Kamann, formerly S. Dreizler) with the integral field spectrograph \textit{Multi Unit Spectroscopic Explorer} (MUSE) at the \textit{Very Large Telescope} (VLT) \citep{kamann2018}. In the wide field mode, MUSE covers a \SI{1}{\arcmin}~x~\SI{1}{\arcmin} field of view (FoV) with a spatial sampling of \SI{0.2}{\arcsec} and a spectral sampling of \SI{1.25}{\angstrom} (resolving power of $1770 < R < 3590$) in the wavelength range from \num{4750} to \SI{9350}{\angstrom}. MUSE allows to extract spectra of some thousand stars per exposure.

The overall aim of our stellar census in \gc{}s with MUSE is twofold.  On the one hand we will investigate the spectroscopic properties of more than half a million of cluster members down to the \ms{}. On the other hand we will investigate the dynamical properties of \gc{}s. Upcoming publications are in the field of multiple populations (Husser et al. submitted, Latour et al. submitted, Kamann et al. in preparation), the search for emission line objects \citep[and Göttgens et al., submitted]{goettgens2019}, cluster dynamics like the overall rotation \citep{kamann2018}, and gas clouds in direction of the clusters \citep{wendt2017}. Since this survey represents the first blind spectroscopic survey of \gc{} cores, we expect many unforeseen discoveries.

\begin{figure}
\resizebox{\hsize}{!}{\includegraphics{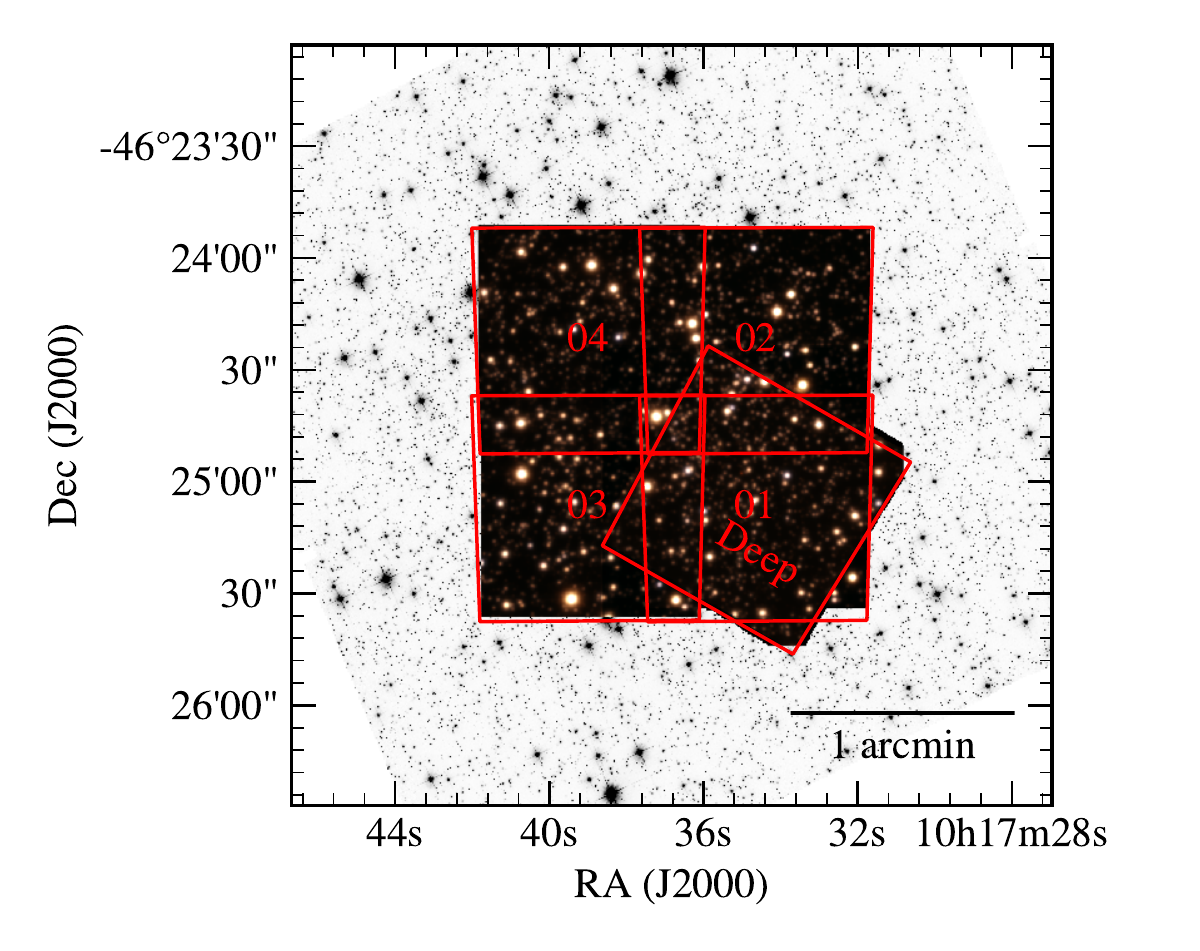}}
\caption{The pointing chart of the MUSE instrument for the \gc{} \ngc{3201}. The inverted background image is from the ACS \gc{} survey \citep{acs1}. The MUSE FoV is a collapsed RGB image created from the spectral cubes. Note the overlapping areas between the pointings.}
\label{fig:pointings}
\end{figure}

For this paper, we used all observations of \ngc{3201} obtained before May 2019. The pointing scheme can be seen in \autoref{fig:pointings}. In addition to the pointings listed in \citet{kamann2018}, we also implemented a deep field mainly to go deeper on the \ms{}. \Autoref{table:observations} lists the number of visits and total integration times for the different pointings. Our data include adaptive optics (AO) observations, our standard observing mode since the AO system was commissioned in October 2017. The AO observations are treated in the same way as our previous observations. The only difference is that a wavelength window of every spectrum around the sodium lines (\SIrange{5805}{5965}{\angstrom}) has to be masked due to the AO laser emission. During each visit, the pointing was observed with three different instrument derotator angles (0, 90, 180\footnote{The deep pointing was observed with an additional derotator angle of 270 degrees.} degrees) in order to reduce systematic effects of the individual MUSE spectrographs.

\begin{figure}
\resizebox{\hsize}{!}{\includegraphics{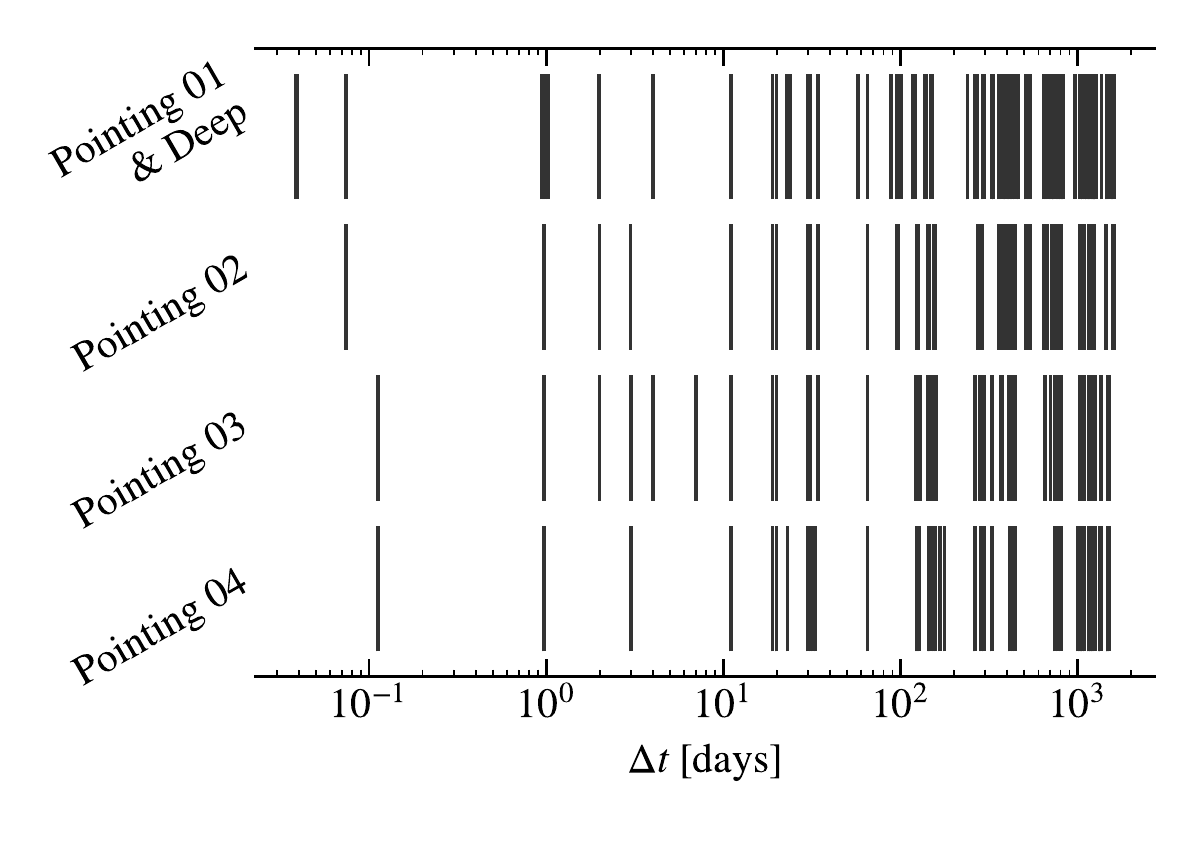}}
\caption{All possible measurement baselines per star and pointing. $\Delta t$ is the time between two observations.}
\label{fig:pointing_epochs}
\end{figure}

\begin{figure}
\resizebox{\hsize}{!}{\includegraphics{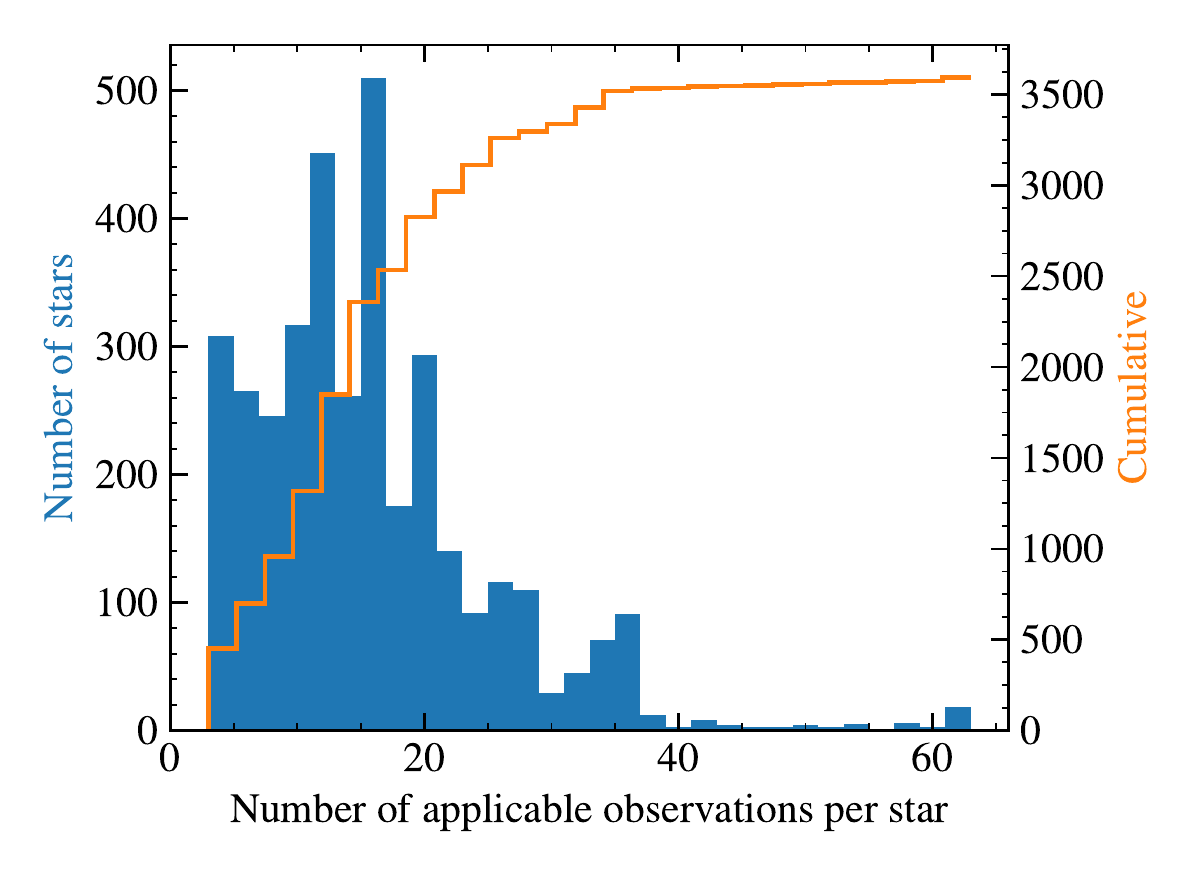}}
\caption{Blue: histogram of applicable (accepted after filtering) number of observations per star. Orange: cumulative distribution of the same data.}
\label{fig:number_of_observations}
\end{figure}

\Autoref{fig:pointings} shows our observation scheme for \ngc{3201}. We secured at least 12 observations per pointing (see \autoref{table:observations} and \autoref{fig:pointing_epochs}). Thanks to the overlapping pointings, up to 63 independent observations per star are available in the central region (see \autoref{fig:number_of_observations}).

To differentiate between radial velocity signals from short period binaries and those of longer period binaries, we observed each of the four pointings twice with a separation of two to three hours in a single night. We already had this distinction sometimes for the stars observed within the same night in the overlapping region of two pointings (like for the published black hole candidate). Without that short baseline a hard binary with a period below one day could mimic the signal of a companion of a black hole candidate at a longer period (temporal aliasing).

The reduction of all exposures was carried out using the standard MUSE pipeline \citep{weilbacher2012,weilbacher2014}.
The extraction from the final product of each visit (epoch) is done by a PSF-fitting technique implemented in the software \pampelmuse{} \citep{kamann}. After an extraction is finished the individual spectra are fitted against a synthetic stellar library to determine stellar properties with \spexxy{} \citep{husser}. The extraction and stellar parameter fitting methods are summarised in \citet{husser2016} and \citet{kamann2018}.

\subsection{Radial velocities}
In this study, we want to investigate if a given star is in a binary system using the radial velocity method. Binaries that are able to survive in the dense environment of a \gc{} are so tight, that even HST is not able to spatially resolve the binary components at the distance of \ngc{3201}. For this reason, the binary system will appear as a single point-like source. Except for the case when both stars have similar brightness (see \autoref{sec:simulation} on how we handle these binaries), the extracted MUSE spectrum will be dominated by one star. The measured radial velocity variation of this star can be used to determine the properties of the binary system indirectly.

\begin{figure}
\resizebox{\hsize}{!}{\includegraphics{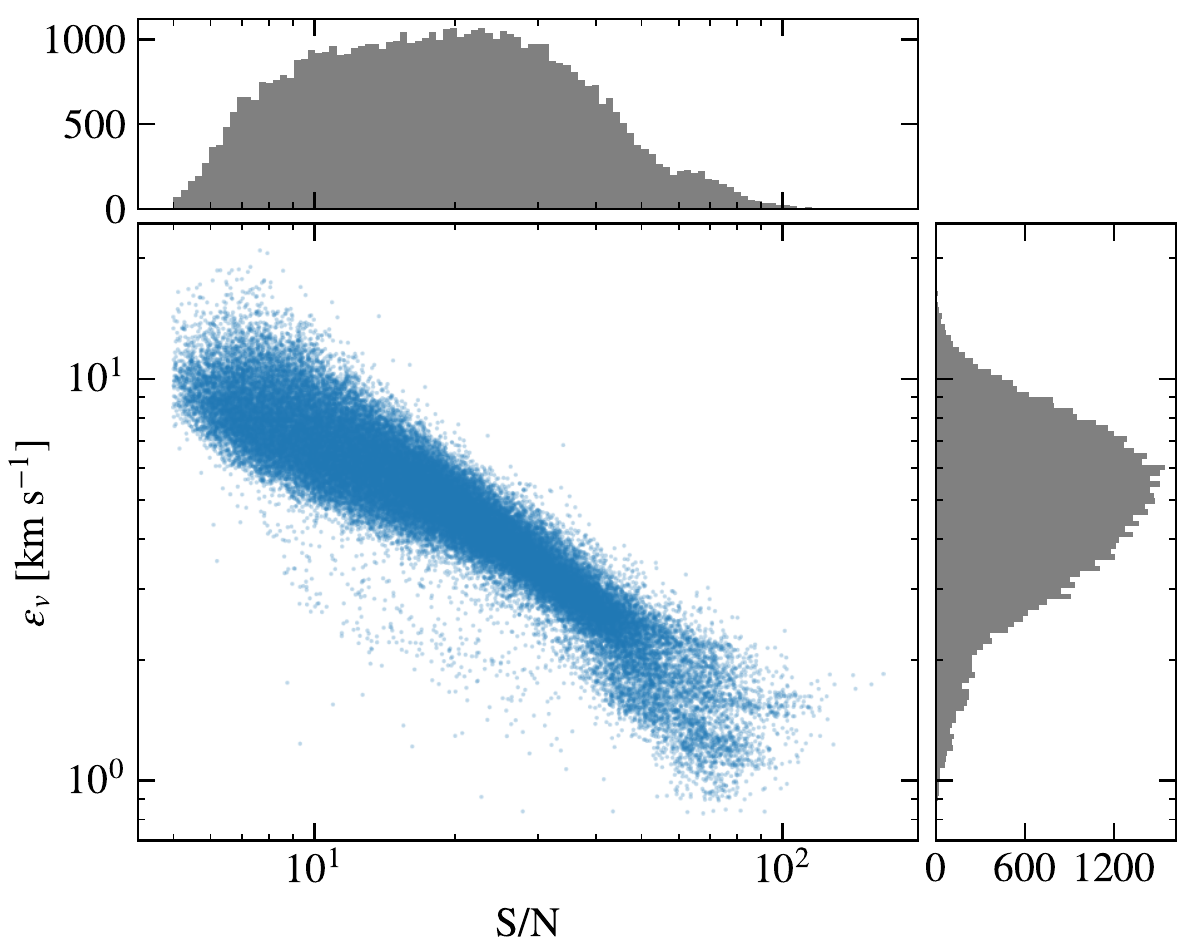}}
\caption{Radial velocity uncertainties $\epsilon_v$ of all spectra in the sample used for \ngc{3201} as a function of the spectral signal-to-noise. %The dashed red line indicates the radial velocity uncertainty limit of \SI{1}{\kilo\meter\per\second}
}
\label{fig:vrad}
\end{figure}

Since we aim for the highest possible radial velocity precision, we checked if the standard wavelength calibration of the MUSE instrument and the uncertainties calculated by the extraction and fitting routines are reliable. For this purpose we analysed the wavelength shift of the telluric lines in the extracted spectra per observation to correct for tiny deviations in the wavelength solution of the pipeline \citep[see Sect.~4.2 in][for details]{kamann2018}.
% with a reasonable signal to noise ($S/N > 30$). The distribution of this wavelength shift is $S/N$ dependent and needs to be neglected by binning among the S/N. These distributions gives us good estimates per S/N bin and observation for the deviation of the wavelength calibration done by the data reduction pipeline and the uncertainties of the spectral fit. All radial velocities per observation are corrected by the mean of the tellurics velocity. Their uncertainties were calculated by the uncertainty propagation of the standard deviation of the tellurics velocity and their fit uncertainty. 
We also cross checked the radial velocity of each spectrum obtained by the full-spectrum fit using a cross-correlation of the spectrum with a PHOENIX-library template \citep{husser} (see criterion in \autoref{sec:finalsample}). \Autoref{fig:vrad} shows the resulting uncertainties $\epsilon_v$ as a function of $S/N$ of the spectra in our final sample of \ngc{3201}. We get reliable radial velocities from $S/N > 5$ spectra \citep{kamann2018}. At $S/N = 10$ the uncertainties are $\epsilon_v \approx \SI{7.0(17)}{\kilo\meter\per\second}$, at $S/N \gtrsim 50$ the uncertainties are $\epsilon_v \approx \SI{1.7(5)}{\kilo\meter\per\second}$. These uncertainties represent our spectroscopic radial velocity precision, the accuracy of the MUSE instrument (wavelength shift between two observations) is below $\SI{1}{\km\per\second}$ \citep{kamann2018}. %\comment{(JB: It seems too me that the absolute uncertainty is less important if the relative is very precise?)}.

\subsection{Selection of the final sample}
\label{sec:finalsample}
For the sample selection two things are crucial. Firstly, we only want reliable radial velocities with Gaussian distributed uncertainties.
Secondly, our sample should only consist of single and multiple star systems which are cluster members, excluding Galactic field stars. Finally, stars that mimic radial velocity variations due to intrinsic pulsations should be excluded.
To ensure this the following filters are applied:
\begin{itemize}
 \item The reliability of every single spectrum is ensured by the following criteria, which are either 0 (false) and 1 (true).
 
 We apply a cut for the signal to noise $S/N$ of a spectrum
 \begin{equation*}
     R_{S/N} = S/N \geq 5 \p{}
 \end{equation*}
 The quality of the cross-correlation is checked using the $r$\hbox{-}statistics and $\operatorname{FWHM}$ defined by \citet{tonry1979}
 \begin{equation*}
     R_\textrm{cc} = r \geq 4 \wedge \operatorname{FWHM} > \SI{10}{\angstrom} \p{}
 \end{equation*}
 A plausible uncertainty $\epsilon_\textrm{v,cc}$ of the cross-correlation radial velocity $v_\textrm{cc}$
 \begin{equation*}
     R_{\epsilon_\textrm{v,cc}} = \epsilon_\textrm{v,cc} > \SI{0.1}{\kilo\meter\per\second} \;,
 \end{equation*}
 and a plausible uncertainty $\epsilon_\textrm{v}$ of the full-spectrum fit radial velocity $v$ 
 \begin{equation*}
    R_{\epsilon_\textrm{v}} = \epsilon_\textrm{v} > \SI{0.1}{\kilo\meter\per\second}
 \end{equation*}
 is ensured. We check that the velocity is within $3 \sigma$ of the cluster velocity $v_\textrm{cluster}$ and the cluster dispersion $\sigma_\textrm{v,cluster}$ including the standard deviation taken from the \citet[][2010 edition]{harris} catalogue. For typical binary amplitudes we add \SI{30}{\kilo\meter\per\second} to the tolerance, yielding
 \begin{equation*}
     \tiny
     R_v = \frac{|v - v_\textrm{cluster}|}{\sqrt{\epsilon_\textrm{v}^2 + \sigma_\textrm{v,cluster}^2 + (\SI{30}{\kilo\meter\per\second})^2}} \leq 3 \p{} % \vee \frac{v}{\sqrt{\epsilon_\textrm{v}^2 + (\SI{10}{\kilo\meter\per\second})^2}} \leq 3
 \end{equation*}
 The last criterion is the check if cross-correlation and full-spectrum fit are compatible within $3 \sigma$ with each other,
 \begin{equation*}
     R_{v = v_\textrm{cc}} = \frac{|v - v_\textrm{cc}|}{\sqrt{\epsilon_\textrm{v}^2 + \epsilon_\textrm{v,cc}^2}} \leq 3 \p{}
 \end{equation*}
 We combine all these criteria in an empirically weighted equation, which gives us a result between 0 and 1,
 \begin{equation*}
    R_\textrm{total} = (2 R_{S/N} + 10 R_\textrm{cc} + R_{\epsilon_\textrm{v,cc}} + 3 R_v + 2     R_{\epsilon_\textrm{v}} + 5 R_{v = v_\textrm{cc}}) / 23 \p{}
 \end{equation*}
 A radial velocity is considered as reliable if the reliability $R_\textrm{total}$ surpasses \SI{80}{\percent}.
 \item We exclude spectra which are extracted within 5 px from the edge of a MUSE cube to avoid systematic effects.
 \item We exclude spectra where \pampelmuse{} was unable to deblend multiple nearby HST/ACS sources (this is of course seeing dependent and affects mostly faint sources).
 \item To identify confusion or extraction problems in a MUSE spectrum, we calculated broad-band magnitudes from the spectrum in the same passband that was used in the extraction process and calculated the differences between input and recovered magnitudes. We accept only spectra with a magnitude accuracy $> \SI{80}{\percent}$ \citep[see Sect.~4.4 in][for more details]{kamann2018}.
 \item In some low S/N cases the cross-correlation and full-spectrum fit returned similar but wrong radial velocities due to noise in the spectrum. These could be outliers by several $\SI{100}{\kilo\meter\per\second}$ introducing a strong variation in the signal and passing the reliability criteria described before. To find outliers we generally compare all radial velocities $v_i$ (with their uncertainties $\epsilon_{v,i}$) of a single star with each other. We define the set of all radial velocities as $A = \{v_i\}$. An outlier $v_x$ in the subset $B = A\,\diagdown\,\{v_x\}$ of all radial velocities except the inspected radial velocity, is identified with
 \begin{equation*}
    \frac{\sigma_A}{\sigma_B} > \kappa \wedge \frac{|v_x - \tilde{B}|}{\epsilon_{v,x} + \sigma_B} > \kappa \quad \textrm{with}\;\kappa = 3 \;.
 \end{equation*}
 $\tilde{B}$ is the median of the radial velocities $v_i$ in the set $B$. $\sigma$ is the standard deviation of the corresponding quantity. Using this equation outliers are identified and excluded from further analyses.
 \item We manually set the binary probability in our sample to 0 for 10 \RRL{}- and 5 \SXP{}-type stars known in the \textit{Catalogue of Variable Stars in Galactic Globular Clusters} \citep{clement2001, clement2017} and \citet{arellano2014}.\footnote{The \RRL{}-type stars are the brightest stars showing radial velocity variations in our sample and we verified that some of them show Keplerian-like signals with periods $< \SI{1}{d}$ and high eccentricities.} 
 %\item We also use a differential photometry method on our MUSE spectra introduced in \citet{giesers} and use the same statistical method described later in \autoref{sec:stats} to get the probability of each star to be photometric variable.
 \item Since \ngc{3201} has a heliocentric radial velocity of \SI{494.0(2)}{\kilo\meter\per\second} and a central velocity dispersion of \SI{5.0(2)}{\kilo\meter\per\second} \citep[][2010 edition]{harris}, we select members by choosing stars with mean radial velocities $> \SI{400}{\kilo\meter\per\second}$.
 \item Our final sample only consists of member stars with at least 3 spectra per star after all filters have been applied.
\end{itemize}

We end up with \num{3553} stars and \num{54883} spectra out of \num{4517} stars and \num{68084} spectra in total. Due to different observational conditions the stars near the confusion limit cannot be extracted reliably from every observation. \Autoref{fig:number_of_observations} shows a histogram of all applicable observations per star for the final sample of \ngc{3201}. We have \num{3285} (\SI{91}{\percent}) stars with five or more epochs, \num{2637} (\SI{73}{\percent}) with ten or more epochs, and \num{918} (\SI{26}{\percent}) stars with 20 or more epochs.

\subsection{Mass estimation}
We estimate the mass of each star photometrically by comparing its colour and magnitude from the ACS \gc{} survey \citep{acs1,acs2} with a PARSEC isochrone \citep{parsec}. For the \gc{} \ngc{3201}, we found the best matching isochrone compared to the whole ACS Colour-magnitude diagram (CMD) with the isochrone parameters
[M/H]~=~\SI{-1.39}{dex} \citep[slightly above the comparable literature value $\lbrack\mathrm{Fe/H}\rbrack$ = \SI{-1.59}{dex}, ][2010 edition]{harris}, 
age~=~\SI{11}{\giga\year}, extinction $E_{B-V}$~=~\num{0.26}, and distance~=~\SI{4.8}{\kilo\parsec}. The mass for each star is determined from its nearest neighbour on the isochrone.
For \bss{}s we assume a mass of \SI{1.20(5)}{\solarmass} \citep{baldwin2016} instead.

\subsection{The electronic catalogue}
An outline of the final sample of the radial velocities is listed in \autoref{table:catalogue}. The columns ACS Id, position (RA, Dec.), and V equivalent magnitude from the ACS catalogue \citep{acs1} are followed by the mean observation date of the combined exposures (BMJD), the barycentric corrected radial velocity $v_\textrm{r}$, and its uncertainty $\epsilon_v$ from our MUSE observations. The last column contains the probability for variability in radial velocity $P(\chi_i^2, \nu_i)$, determined as described in \autoref{sec:stats}. Three versions of the catalogue, one with the full final sample (see \autoref{sec:finalsample}), one with non-member stars, and one with the \RRL{} and \SXP{} stars are available at the CDS and on the project website \url{https://musegc.uni-goettingen.de/}.

\section{\mocca{} simulation of NGC 3201}
\label{sec:simulation}
We want to compare our observations of binaries in \ngc{3201} with models. Our first attempt using a toy model with realistic binary parameter distributions for this specific cluster like period, eccentricity, mass ratio, etc. failed since the creation of binaries from these distributions is unrealistic. \citet{moe2017} demonstrated that randomly drawing values of these orbital parameter distributions does not lead to realistic currently observed binary distributions even when the parameter distributions themselves are realistic. Especially not for \gc{}s where dynamical interactions between stars play an important role. Thus we have to use more sophisticated simulations of \gc{}s including all relevant physics for the cluster dynamics and the binary evolution. Simulating the dynamical evolution of realistic \gc{}s in a Galactic potential up to a Hubble time is computationally expensive. %\citet{rodriguez2016} summarised the recovery of observational properties of \gc{}s using direct N-body or statistical approaches.

\begin{figure}
\resizebox{\hsize}{!}{\includegraphics{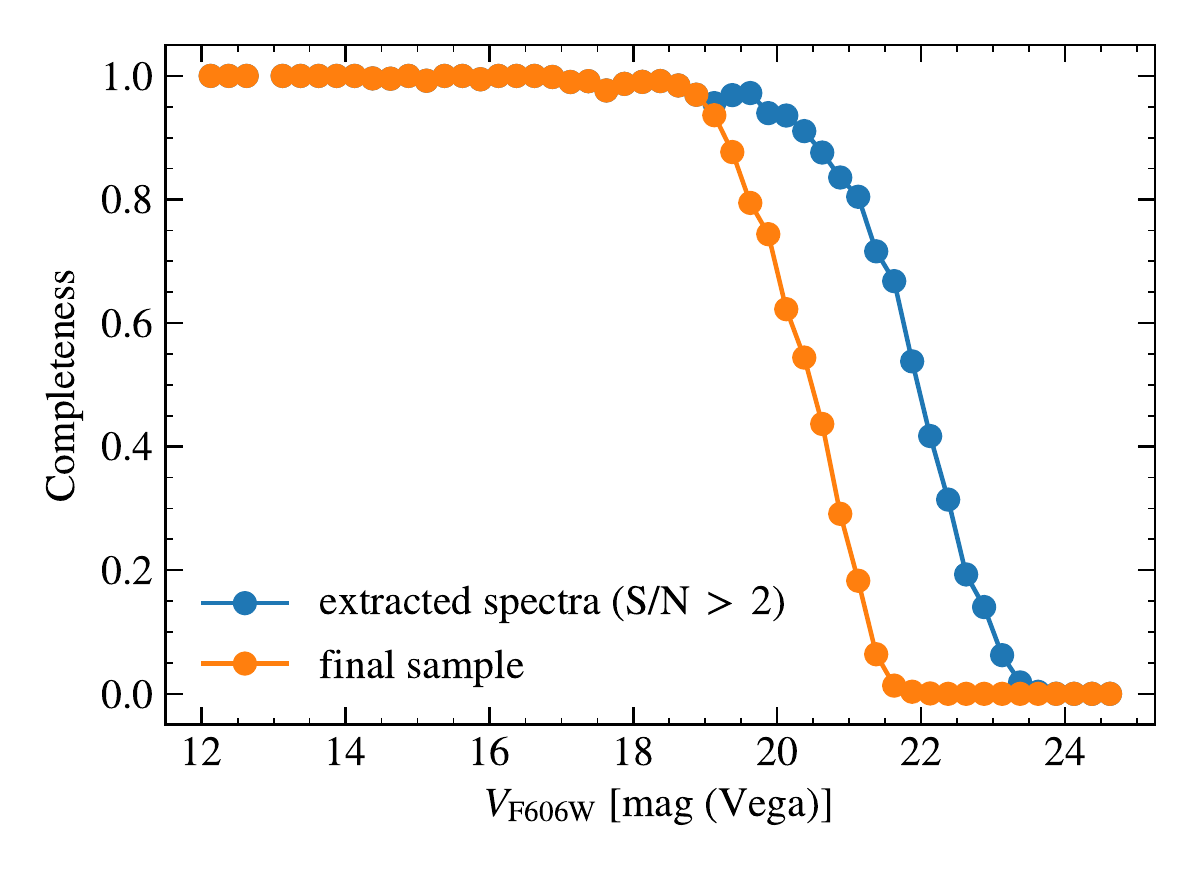}}
\caption{The completeness as a function of magnitude for all MUSE observations of \ngc{3201}, using photometry from the ACS \gc{} survey \citep{acs1}. The fraction of all possible extracted spectra to observed spectra within the MUSE FoV is shown. Blue: extracted spectra with S/N~>~2. Orange: extracted spectra in the final sample (see \autoref{sec:finalsample}). (The uncertainties are smaller than the points.)}
\label{fig:completeness}
\end{figure}

\begin{table}
 \caption{\Gc{} properties for the \mocca{} model in comparison to observed values.}
 \label{table:ngc3201}
 \centering
 \tiny
 \begin{tabular}{r c c}
 \hline\hline
 RA & \multicolumn{2}{c}{10\nfh17\nfm36\fs82 (a)} \\
 Dec. & \multicolumn{2}{c}{-46\degr 24\arcmin 44\farcs9 (a)} \\
 Distance to Sun \si{\kilo\parsec} &
 \multicolumn{2}{c}{\SI{4.9}{\kilo\parsec} (a)} \\
 Metallicity [Fe/H] & \multicolumn{2}{c}{\SI{-1.59}{dex}} \\
 Barycentric radial velocity & \multicolumn{2}{c}{\SI{494.0(2)}{\kilo\meter\per\second} (a)} \\
 \hline\hline
  & MOCCA & Literature \\    % table heading
 \hline     
 Central velocity dispersion [\si{\kilo\meter\per\second}] & \num{5.23} (*) & \num{5.0(2)} (a) \\
 Core radius [\si{\parsec}] & \num{1.54} & \num{1.85} (a) \\
 Half-light radius [\si{\parsec}] & \num{4.16} & \num{4.42} (a) \\
 Total V-band luminosity [$L_\odot$] & \num{9.27e4} & \num{8.17e4} (a) \\
 Central surface brightness [\si{L_{\odot}\per\parsec\squared}] & \num{2.25e3} & \num{9.14e2} (a) \\
 Age [Gyr] & \num{12} & \num{12.0(8)} (b) \\
 Mass [\si{\solarmass}] & \num{2.3e5} & \num{1.49(9)e5} (c) \\
 Total binary fraction [\si{\percent}] & \num{8.72} & --- \\
 \hline
 \end{tabular}
 \tablefoot{(a)~\citet[][2010 edition]{harris} catalogue; (b)~\citet{dotter2010}; (c)~\citet{baumgardt2018}; (*)~computed for stars brighter than 3 mags below \ms{} turn-off).}
\end{table}

\subsection{The model}
An approach that has been used to extensively simulate realistic \gc{} models is the Monte Carlo method for stellar dynamics that was first developed by \citet{1971Ap&SS..14..151H}. This method combines the statistical treatment of relaxation with the particle based approach of direct \textit{N}-body methods to simulate the long term evolution of spherically symmetric star clusters. The particle approach of this method enables the implementation of additional physical processes (including stellar/binary evolution) and is much faster than direct \textit{N}-body methods. In recent years, an extensive number of \gc{} models have been simulated with the \mocca{} (\textit{MOnte Carlo Cluster simulAtor}) \citep{1998MNRAS.298.1239G,2013MNRAS.429.1221H,2013MNRAS.431.2184G,2017MNRAS.464L..36A,2018MNRAS.480.5645H,belloni2019} and CMC (\textit{Cluster Monte Carlo}) codes \citep{2000ApJ...540..969J,2013ApJS..204...15P,2015ApJ...800....9M,2016PhRvD..93h4029R,2017ApJ...834...68C}.

For the purpose of this project we needed a simulated \gc{} model that would have present-day observational properties comparable to \ngc{3201} and that would also contain stellar-mass black holes like the one we found \citep{giesers}. Therefore, we used results from the MOCCA Survey Database I \citep{2017MNRAS.464L..36A} to identify initial conditions for a model that had global properties close to \ngc{3201} at an age of 12\,Gyr. This model has been subsequently refined with additional \mocca{} simulations based on slightly modified initial conditions to find an even better agreement with the properties of \ngc{3201} at an age of 12\,Gyr including total luminosity, core and half-light radii, central surface brightness, velocity dispersion and binary fraction.

The \mocca{} code utilizes the Monte Carlo method to compute relaxation. Additionally, it uses the fewbody code \citep[small number N-body integrator]{fregeau2004} to compute the outcome of strong interactions involving binary stars. \mocca{} also makes use of the SSE \citep{2000MNRAS.315..543H} and BSE \citep{2002MNRAS.329..897H} codes to carry out stellar and binary evolution of all stars in the cluster. \mocca{} results have been extensively compared with results from direct \textit{N}-body simulations and they show good agreement for both the evolution of global properties and number of specific objects \citep{2013MNRAS.431.2184G,2016MNRAS.458.1450W}.

The following initial parameters were used for the \mocca{} model we simulated specifically for this work to have a comparison to \ngc{3201}: The initial number of objects was set to $N=5.5\times 10^{5}$. The initial binary fraction was set to \SI{50}{\percent}, resulting in \num{2.25e5} single stars and \num{5.5e5} stars in binary systems, thus in total \num{8.25e5} stars in the initial model. The initial binary properties were set as described in \citet{2017MNRAS.471.2812B}. The metallicity of the model was set to ($Z=5.1 \times 10^{-4}$) to correspond to the observed metallicity of \ngc{3201} provided in the \citet[][2010 edition]{harris} catalogue. We used a \citet{1966AJ.....71...64K} model with a central concentration parameter ($W_{0}$) of 6.5. The model had an initial half-mass radius of \SI{1.65}{\parsec} and a tidal radius of \SI{165}{\parsec}. The initial stellar masses were sampled using a two-component \citet{kroupa2001} IMF (initial mass function) with masses in the range $0.08 M_{\odot}$ to $100 M_{\odot}$. Neutron star natal kicks were drawn from the \citet{2005MNRAS.360..974H} distribution. For black holes, natal kicks were modified according to black hole masses that were computed following \citet{2002ApJ...572..407B}. The properties of the evolved simulated model at \SI{12}{Gyr} are listed in \autoref{table:ngc3201} and compared to literature values. The \mocca{} model for \ngc{3201} is able to well reproduce the observational core radius, half-light radius, central velocity dispersion, and the V-band luminosity provided in \citet[][2010 edition]{harris}. The central surface brightness for the \mocca{} model is a factor of 2.5 larger than the observed one which is unfortunately not given with an uncertainty. In contrast, \citet{McLaughlin2005}, for example, found about twice the observational central surface brightness listed in \citet[][2010 edition]{harris}. Given the observational uncertainties and limitations, the agreement between the simulated model and the observed values are reasonable.

\subsection{The mock observation}
We project the simulation snapshot at \SI{12}{Gyr} in Cartesian coordinates using the \textsc{COCOA} code \citep{askar2018a}. We use the distance modulus of \ngc{3201} at a distance of  \SI{4.9}{\kilo\parsec} \citep[][2010 edition]{harris} to get the apparent magnitude of all stars. For binary stars we combine the two components into a single magnitude. We select the stars within the MUSE FoV (of all pointings). Additionally we use our observational completeness function for \ngc{3201} (see \autoref{fig:completeness}) to select stars accordingly.

After we identified a set of stars comparable to the MUSE observations, we created mock radial velocities for these stars. To this aim, we let our observational data guide the simulation. For every observed star (in our MUSE sample) we randomly picked a \mocca{} star with comparable magnitude. We then used the epochs of the observed star to create new radial velocity measurements. For single stars we simply assumed a radial velocity of \SI{0}{\kilo\meter\per\second} for every time step. For binaries we calculated the radial velocity for the brighter component from a Keplerian orbit using the \mocca{} properties of the binary system (masses, period, inclination, eccentricity, argument of periastron, and periastron time). However, due to the spectral resolution of MUSE its measured radial velocity amplitude will be influenced by the flux ratio of the two components of the binary system. From Monte Carlo simulations we found that the theoretical expected radial velocities $v_\mathrm{r,theoretical}$ are linearly damped with the flux ratio:

\begin{equation}
v_\mathrm{r,measured} = v_\mathrm{r,barycentre} + \left(1 - \frac{f_2}{f_1}\right) \, v_\mathrm{r,theoretical}\;.
\end{equation}

With the flux of the brighter component $f_1$ and the fainter component $f_2$. $v_\mathrm{r,barycentre}$ is the radial velocity of the barycentre, which can be neglected if only relative velocities are of interest. In the case of both components having the same flux, the measured radial velocity $v_\mathrm{r,measured}$ amplitude will be zero. Thus, for example, twin \ms{} binary stars are hard to find. In the case when the fluxes of the components are very different, the observed radial velocity amplitude corresponds to the amplitude of the brighter component. We apply this damping for every simulated radial velocity. Finally we scatter these radial velocities within the uncertainties taken from the MUSE corresponding measurements.

\section{The binary fraction of \ngc{3201}}
\label{sec:stats}
\Autoref{sec:simulation} reflects the efforts required to create realistic models for \gc{}s. We therefore conceived a model independent approach to derive the binary frequency of a \gc{} as well as the probability of an individual star to be radial velocity variable.

\subsection{A new method to detect variable stars}
We here introduce a general statistical method which could be applied to any inhomogeneous sample with a varying number of (few) measurements and a large range of uncertainties. Here it is applied to radial velocity measurements: To detect radial velocity variations in a given star with $m$ measurements, we compute the $\chi_i^2$ for the set of measurements $x_j$ with uncertainties $\sigma_j$
to determine how compatible they are with the constant weighted mean $\overline{x}$ (null hypothesis) of the measurements:
\begin{equation}
 \chi_i^2 = \sum_{j=1}^m \frac{1}{\sigma_j^2} \left( x_j - \overline{x} \right)^2 \textrm{\quad with \quad} \overline{x} = \frac{\sum_{j=1}^m \frac{x_j}{\sigma_j^2}}{\sum_{j=1}^m \frac{1}{\sigma_j^2}} \;. \label{eq:chi2}
\end{equation}

\begin{figure}
\resizebox{\hsize}{!}{\includegraphics{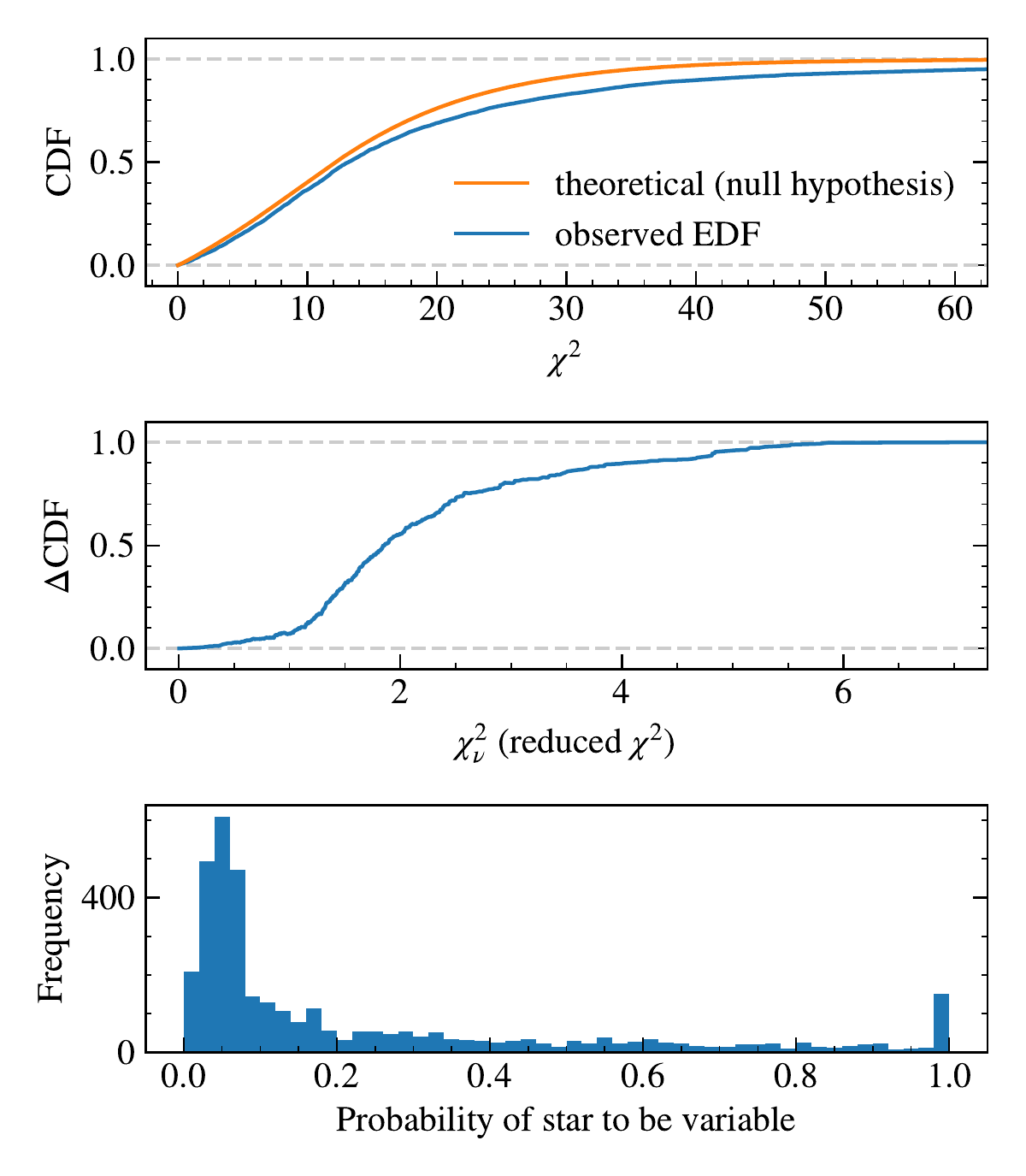}}
\caption{Top panel: superposition of observed \textit{empirical distribution function} (EDF) and expected theoretical \textit{cumulative distribution function} (CDF) using the null hypothesis that no variable star exist in sample. Middle panel: resulting $\Delta$CDF. Bottom panel: histogram of variability probabilities in the sample.}
\label{fig:cdf2pdf}
\end{figure}

A star that shows radial velocity variations larger than the associated uncertainties will have \textit{reduced} $\chi^2 > 1$ on average. In contrast, a star without significant variations will have \textit{reduced} $\chi^2 \approx 1.$
As described in \autoref{sec:observations} each star in our sample has its own number of observations and therefore its own number of degrees of freedom ($\nu = m - 1$, see \autoref{fig:number_of_observations}). 
Assuming Gaussian distributed uncertainties for all measurements
and a common degree of freedom for all stars, we know the expected $\chi^2$-distribution in form of the \textit{cumulative distribution function} (CDF) $F$ for the null hypothesis that all stars show constant signals. 
If there are binary stars with radial velocity variations in our sample, there will be an excess of larger $\chi^2$ values compared to the null hypothesis. 
Therefore, the empirical CDF computed from the measured $\chi^2$ will increase slower than the CDF computed using the null hypothesis.

For a chosen degree of freedom (e.g. $\nu = 3$) we calculate the probability $\bar{P}$ of each star to match 
the null hypothesis through the comparison of the observed \textit{empirical distribution function} (EDF) $F_\textrm{observed}$ with the expected CDF (comparable to the top panel in \autoref{fig:cdf2pdf}):
\begin{equation}
 \bar{P}(\chi_i^2, \nu_i = 3) = \frac{1 - F(\chi_i^2, 3)_\textrm{theoretical}}{1 - F(\chi_i^2, 3)_\textrm{observed}} \;.
\end{equation}
Since the EDF $F_\textrm{observed}$ is simply the fraction of stars below a given $\chi^2$, 
we divide the number of stars we measured below a given $\chi^2$ with the expected number from the known CDF to calculate $P$ (comparable to the middle panel in \autoref{fig:cdf2pdf}).
The probability of a star to be variable for a given degree of freedom $\nu$ is $P = 1 - \bar{P}$, i.e.
\begin{equation}
 P(\chi_i^2, \nu_i) = \frac{F(\chi_i^2, \nu_i)_\textrm{theoretical} - F(\chi_i^2, \nu_i)_\textrm{observed}}{1 - F(\chi_i^2, \nu_i)_\textrm{observed}} \;.
 \label{eq:prob}
\end{equation}
To use the statistical power of the whole sample with the total number of stars $n$, this equation can be generalised to multiple degrees of freedom by noticing that $n \, F_\textrm{theoretical}$ is the expected number of stars below a given $\chi^2$.
The (expected or observed) total number of stars below a given $\chi^2$, taking into account all available degrees of freedom, can be expressed using the number of the stars per degree of freedom $n_\nu$ and adding up the contributions from all degrees of freedom in a 'super CDF':
\begin{equation}
 \widehat{S}(\chi_i^2) = \frac{1}{n} \sum_{\nu}{F(\chi_i^2, \nu)\,n_\nu} \;,
 \label{eq:cdf}
\end{equation}
from which $n \, \widehat{S}$ is the analogue of $n \, F_\textrm{theoretical}$ (and $n \, F_\textrm{observed}$) for multiple degrees of freedom. 
Inserting this into Eq.~\ref{eq:prob} yields
\begin{equation}
 P(\chi_i^2, \nu_i) = \frac{n\,\widehat{S}(\chi_k^2) - k}{n - k} \quad \textrm{with}\;k=\left\vert\left\{\chi^2 \vert \chi^2/\nu < \chi_i^2/\nu_i\right\}\right\vert
 \label{eq:pdf}
\end{equation}
% is the k-th smallest
where $k$ is the number of stars with a \textit{reduced} $\chi^2$ lower than the one of the given star, i. e. $\chi_i^2/\nu_i$ and $\chi_k^2$ .

The upper panel of \autoref{fig:cdf2pdf} shows the observed EDF and the theoretical 'super CDF' $\widehat{S}$. Both functions deviate for increasing $\chi^2$, which indicates that stars in the sample not only show statistical variations. The middle panel shows the resulting $\Delta$CDF (\autoref{eq:pdf}) for our sample. If we evaluate it with the \textit{reduced} $\chi^2_\nu$ we get the probability for every star to vary in radial velocity. The bottom panel of \autoref{fig:cdf2pdf} shows the resulting binary probability distribution of all \ngc{3201} stars in our sample.

Comparing the number of stars with $P > 0.5$ to the total number of stars yields the discovery binary fraction. We checked this approach using Monte Carlo simulations for different star samples of single and binary stars. It should be noted that at this \SI{50}{\percent} threshold there is a balance of false-positives and false-negatives, but that ensures a robust measurement for the overall statistics. Of course, for individual stars the acceptance probability to be a binary should be higher.

The statistical uncertainty of the binary fraction is calculated by the quadratic propagation of the uncertainty determined by bootstrapping (random sampling with replacement) the sample and the difference of the fraction for $P > 0.45$ and $P > 0.55$ divided by 2 as a proxy for the discriminability uncertainty between binary and single stars. %\comment{(JB: why this range?)}

\subsection{Verification of the method on the MOCCA mock observation}
\label{subsec:mock}
\begin{figure}
\resizebox{\hsize}{!}{\includegraphics{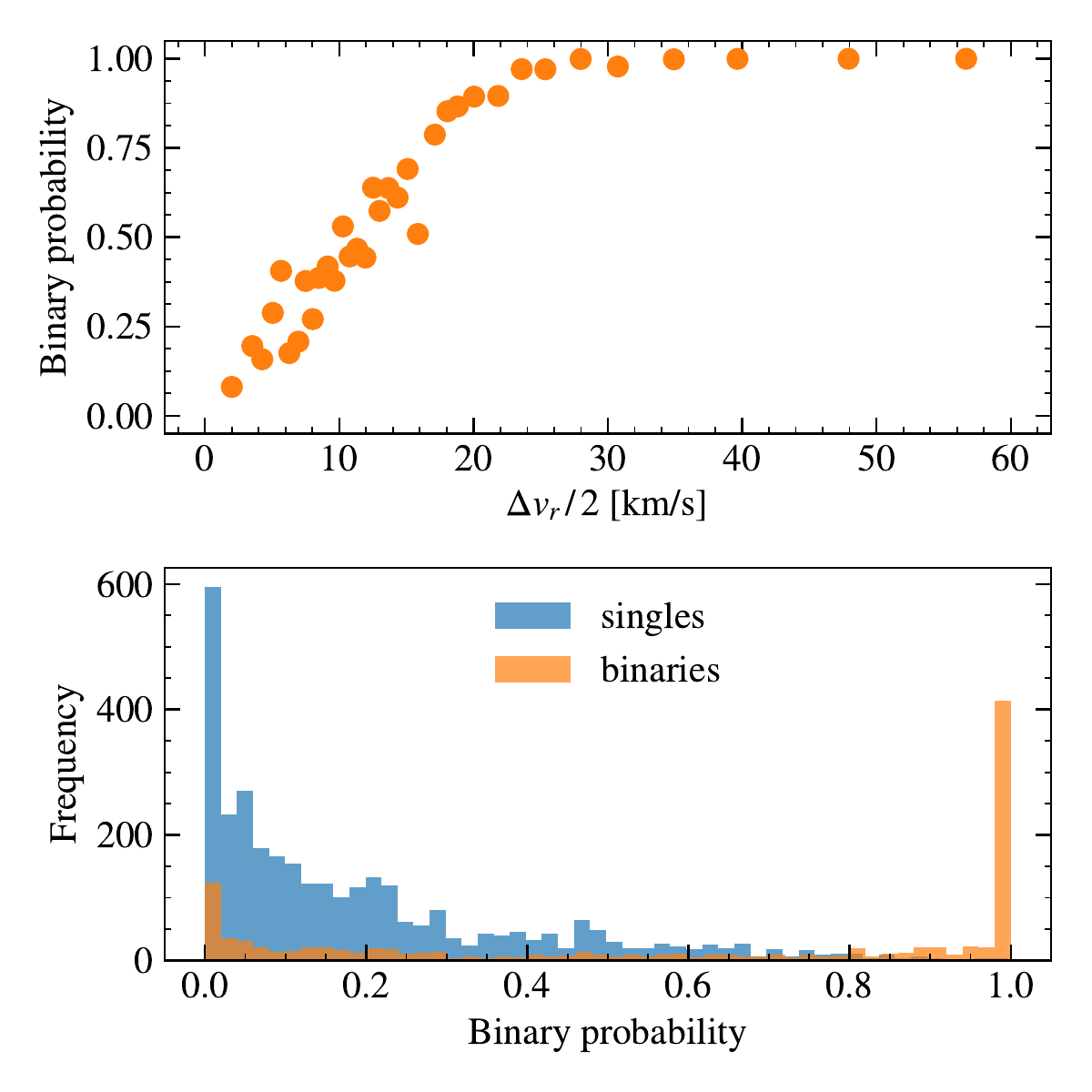}}
\caption{Top panel: the variability probability as a result of the statistical method for all known binaries in the \mocca{} mock observation as a function of $\Delta v_r\,/\,2$, a proxy for the projected radial velocity semi-amplitude. Bottom panel: histograms of the variability probability for known single and binary stars in the MOCCA mock observation.}
\label{fig:sanity}
\end{figure}

To verify the validity of this statistical method, we applied it to the \mocca{} mock observation introduced in \autoref{sec:simulation}. For each star we calculated the variability probability according to \autoref{eq:pdf}. We also calculated a kind of semi-amplitude by bisection of the peak to peak of the simulated radial velocities ($\Delta v_r\,/\,2$) of each star. In the top panel of \autoref{fig:sanity} the mean binary probability in relation to this semi-amplitude for all known binaries in the mock observation is presented, using a binning of 30 stars per data point. In view of the radial velocity uncertainties (see \autoref{fig:vrad}) the statistical method gives plausible results: for example \SI{50}{\percent} of all stars with a semi-amplitude around \SI{10}{\kilo\meter\per\second} are recovered. The bottom panel of \autoref{fig:sanity} shows histograms of the binary probability of the known single stars (in blue) and the known binary stars (in orange). We get a separation of the binary stars (peak at probability $\approx 1$) and single stars (peak at probability $\approx 0$). It shows the power of our approach, as it results in a binary distribution peaked at high probabilities and a single star distribution peaked at low probabilities. Using a threshold of P=0.5 (0.8) would result in a false-positive rate of 324/781 (44/617) stars. These rates are in agreement with the expectations, since the method gives us a balance of false-positives and false-negatives at the \SI{50}{\percent} threshold, as mentioned before.

The statistical method yields a discovery fraction of \SI{22.1(18)}{\percent} for this mock observation. Compared with the true binary fraction in this mock observation of \SI{25.9}{\percent} we get a discovery efficiency of \SI{85.2}{\percent}. This implies that we can expect to detect the vast majority of binary stars present in our observed MUSE sample.

\subsection{Application to the MUSE observations}
Using the statistical approach we obtained an observational binary frequency (discovery fraction) of \SI{17.1(19)}{\percent} within our MUSE FoV. The bottom panel of \autoref{fig:cdf2pdf} shows the resulting binary probability of our stars in the sample. To determine the true binary frequency of the \gc{} \ngc{3201} we have to overcome our observational biases. Our survey is a blind survey and we do not select our targets as in previous spectroscopic studies. Nevertheless, we still have a magnitude limit below which we cannot derive reliable radial velocities and we are confined to the MUSE FoV. In \autoref{fig:completeness} the magnitude completeness of MUSE stars compared to the ACS \gc{} survey \citep{acs1} of \ngc{3201} is shown (blue curve). After filtering (see \autoref{sec:observations}), we end up with the magnitude completeness of the binary survey (orange curve). To get the binary frequency we have to divide the discovery fraction by the discovery efficiency. As described in \autoref{sec:simulation}, the discovery efficiency cannot be derived by simple assumptions (except for inclination of the binary system). Moreover, the efficiency strongly depends on the number of observations and the sampling per star. \Autoref{fig:number_of_observations} shows that our distribution of applicable observations is inhomogeneous and likewise is our time sampling (see \autoref{fig:pointing_epochs}).

We use the \mocca{} mock observation to overcome these biases. Assuming the \mocca{} simulation is realistic and comparable to our MUSE observation, we found a binary frequency of \obf{} within our MUSE FoV, taking the discovery efficiency of \autoref{subsec:mock} into account. We can now use the original \mocca{} simulation and compare its true binary frequency of \SI{8.72}{\percent} with the true binary frequency of the mock observation \SI{25.9}{\percent} and get a factor of \num{0.336}. Applying this factor and the discovery efficiency it is possible to translate the MUSE discovery fraction into the total binary fraction of \ngc{3201} to %\comment{(JB: Does the simulations have a limit on primary mass or luminosity?)}
\begin{equation*}
\large
\cbf{} \p{}
\end{equation*}

The main reason for this value being significantly lower than our observed binary frequency is the central increase of binaries due to mass segregation. Our value is in reasonable agreement with the binary fraction of the \mocca{} simulation, \SI{8.72}{\percent}, considering the assumptions that have to be made in Monte Carlo models. This highlights again that the \mocca{} simulation that we selected represents an accurate model for \ngc{3201}.

We also verified to what extent our study is consistent with the results of \citet{milone2012}, who reported a core binary frequency of \SI{12.8(8)}{\percent}. Without applying any selection function, the \mocca{} simulation yields a binary fraction of \SI{12.5}{\percent} within the core radius of \ngc{3201}. Hence our results appear to be consistent with the study of \citet{milone2012} and the apparent differences (with respect to our discovery fraction of \obf{}) can be attributed to selection effects.

\subsection{Primordial binaries}
We created different \mocca{} simulations of \ngc{3201} to get a comparable mock observation and thus binary fraction to our observations. The best matching \mocca{} simulation (see \autoref{sec:simulation}) has an initial binary fraction of \SI{50}{\percent} which indicates that a significant fraction of primordial binaries is necessary to reproduce the current observations of \ngc{3201}.

Recent studies have shown that populations of cataclysmic variables and X-ray sources observed in several \gc{}s can be better reproduced with \gc{} simulations that assume high primordial binary fractions \citep[e.g.][]{sandoval2017,cheng2018,belloni2019}.
In addition, \citet{leigh2015} and \citet{cheng2019} found that a high primordial binary fraction is also necessary to reproduce the binary fraction outside the half-mass radius and the mass segregation we observe in Galactic \gc{}s.

\begin{figure}
\resizebox{\hsize}{!}{\includegraphics{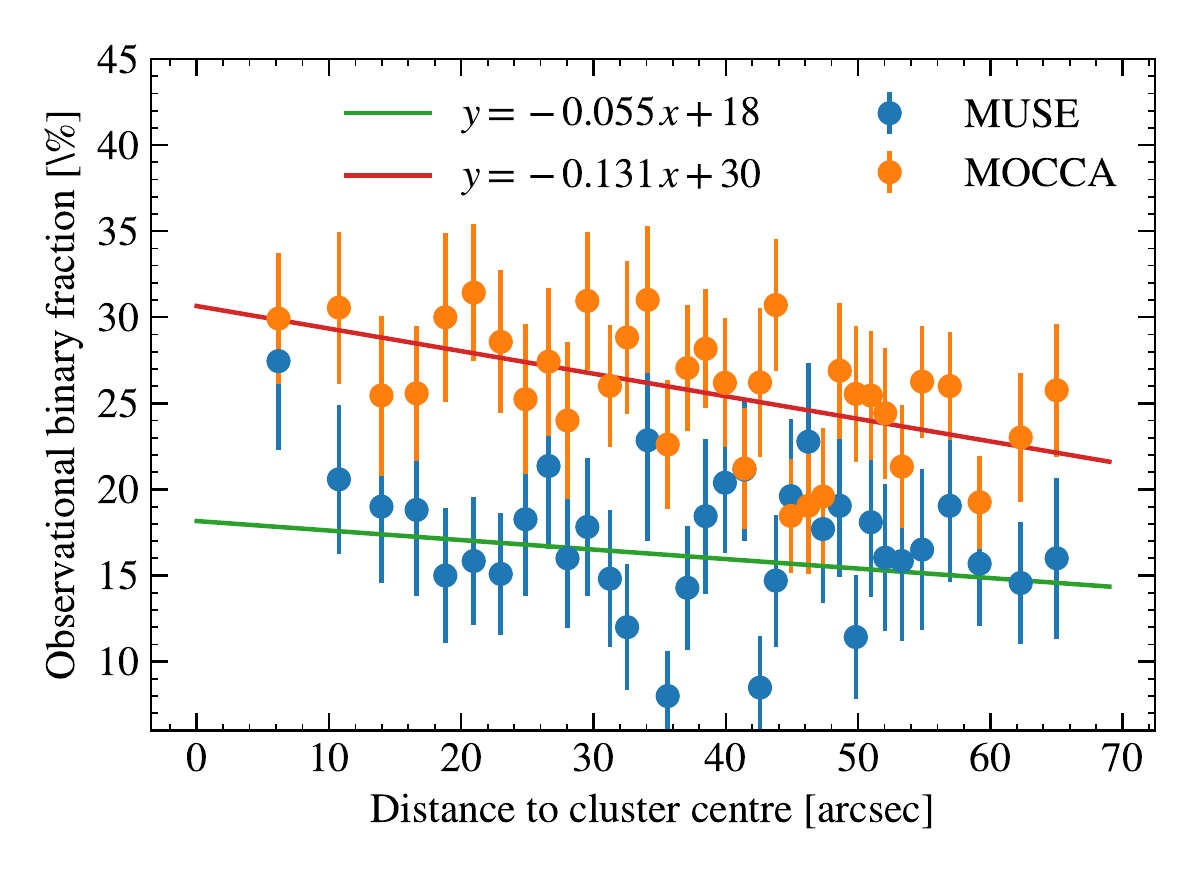}}
\caption{The observational binary fraction of \ngc{3201} in radial bins relative to the cluster centre. Blue: MUSE bins with 100 stars per bin. Orange: \mocca{} MUSE equivalent mock observation, same bins as MUSE.}
\label{fig:radial_fraction}
\end{figure}

\subsection{Mass segregation}
Due to mass segregation it is expected that binary systems (which are in total more massive than single stars) will migrate towards the centre of a \gc{} (see \autoref{sec:introduction}). Thus the binary fraction should increase towards the cluster centre. In \autoref{fig:radial_fraction} the observational binary discovery fraction in relation to the projected distance to the centre of the \gc{} is shown. Since we do not have true distances to the cluster centre, any observed radial trend is expected to be somewhat weaker than the underlying trend with true radii. Nevertheless, we found a mild increase of the binary fraction towards the cluster centre. The slope of a simple linear fit is \SI{-0.055(3)}{\percent} per arcsec with a Pearson correlation coefficient of $-0.28$.
Despite the observational biases and the limited radial distance to the cluster centre this result agrees with the theoretical expectations. Compared to the mock simulation, for which the same projected radial bins have been applied, the result is qualitatively the same. We like to stress that the original \mocca{} simulation has a clear radial trend in the binary frequency (using three dimensional radial bins). %\changed{and a total binary fraction of \SI{20}{\percent} within a central volume with a radius of \SI{0.5}{\parsec}.}

%We get for example 189 (126) stars with a binary probability of \SI{90}{\percent} ($\SI{99.73}{\percent} = 3 \sigma$).

\begin{figure*}
\resizebox{\hsize}{!}{\includegraphics{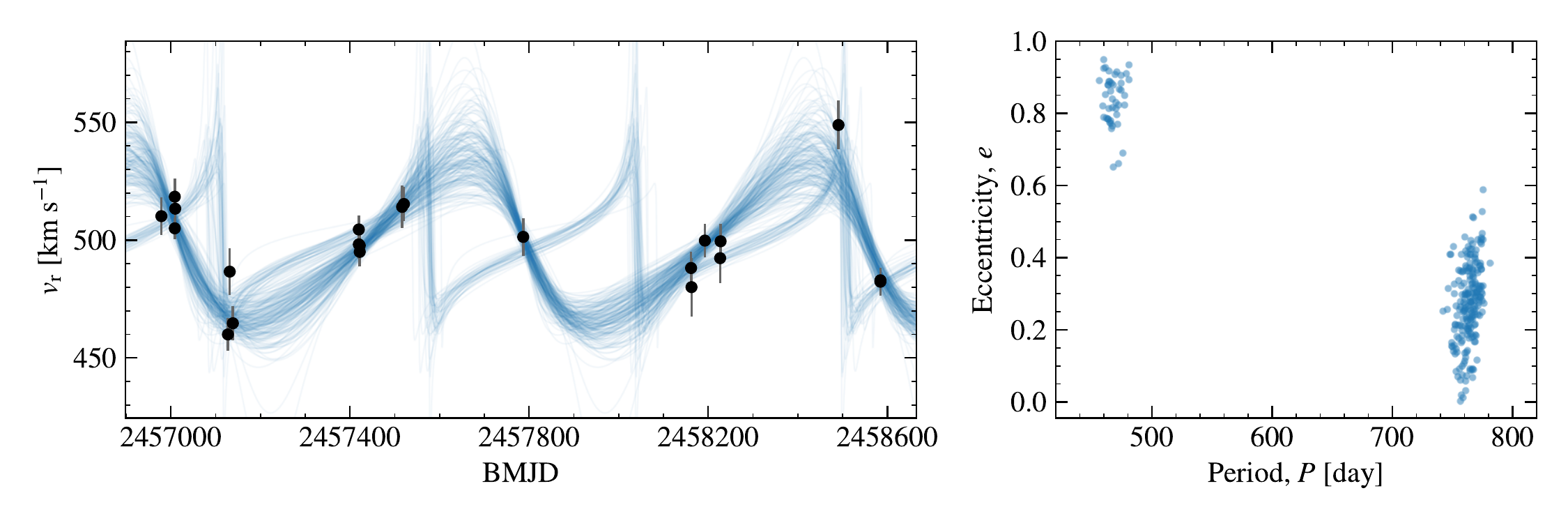}}
\caption{Left: radial velocities (black data points) of the companion of the black hole candidate with ACS Id \#5132. The blue curves are the possible orbital solutions determined by \joker{} for this star. Right: period-eccentricity plot of these samples.}
\label{fig:joker_example}
\end{figure*}

\section{Determination of orbital parameters}
\label{sec:orbits}
Constraining the orbital properties (six standard Keplerian parameters) of a binary system with few or imprecise radial velocity measurements is difficult. Naively one could assume that in dense \gc{}s only hard binaries (with high binding energy and typically short period) could survive and should circulise their orbit during their lifetime. However, not only due to dynamical interactions, a significant number of binaries are expected to be on eccentric orbits \citep{hut1992}. Methods like the generalised Lomb-Scargle periodogram (GLS) \citep{zechmeister} are good to find periods in unequally sampled data. GLS is good for finding periods in circular orbits, but tends to fail in eccentric configurations. The APOGEE team also faced the same challenges and created a tool named '\joker{}' \citep{joker}.

\subsection{\joker{}}
\label{sec:joker}

\joker{} is a custom Monte Carlo sampler for sparse or noisy radial velocity measurements of two-body systems and can produce posterior samples for orbital parameters even when the likelihood function is poorly behaved.

We follow the method of \citet{apogee} to find orbital solutions for our \ngc{3201} stars. Our assumptions to use \joker{} are accordingly:
\begin{itemize}
 \item \textbf{Two bodies}: We assume the multiple star systems in \ngc{3201} to consist of two stars or only two stars to be on the dominant timescale for the dynamics of the system (hierarchical multi-star system).
 \item \textbf{SB1}: The light of the binary components adds up in the MUSE spectrograph as a single source. We assume one of the two binary components does not contribute significantly to the spectrum (single-lined spectroscopic binary = SB1). In the case both components have similar magnitudes (double-lined spectroscopic binary = SB2) it is not possible for us to separate the components for typical radial velocity amplitudes, due to the low spectral resolution of MUSE. As mentioned in \autoref{sec:simulation}, the amplitude of SB2 systems depends on the flux ratio of the two binary components.
 \item \textbf{Isolated Keplerian systems}: The radial velocity variations are due to orbital motions and not caused by possible intrinsic variations (e.g. pulsations). Moreover, the cluster dynamics happens on a larger timescale than the orbital motion in the binary system.
 \item \textbf{Gaussian distributed uncertainties}: We ensured all radial velocity uncertainties to be free from systematic effects, independent from each other and represent purely Gaussian noise \citep[for details see][and \autoref{sec:finalsample}]{kamann2018}.
\end{itemize}

We generated $2^{29} = \num{536870912}$ prior samples for the period range \SI{0.3}{\day} to \SI{4096}{\day}. We requested 256 posterior samples for each star with a minimum of 5 observations and a minimum \SI{50}{\percent} probability to be a variable star. In our sample \num{515} met these conditions and could be processed with \joker{}. If fewer than 200 posterior samples were found for one star, we used a dedicated MCMC run as described in \citet{joker} to get 256 new samples. We analysed their posterior distributions and will focus on objects with unimodal or bimodal distributions in the following.

\begin{figure*}
\resizebox{\hsize}{!}{\includegraphics{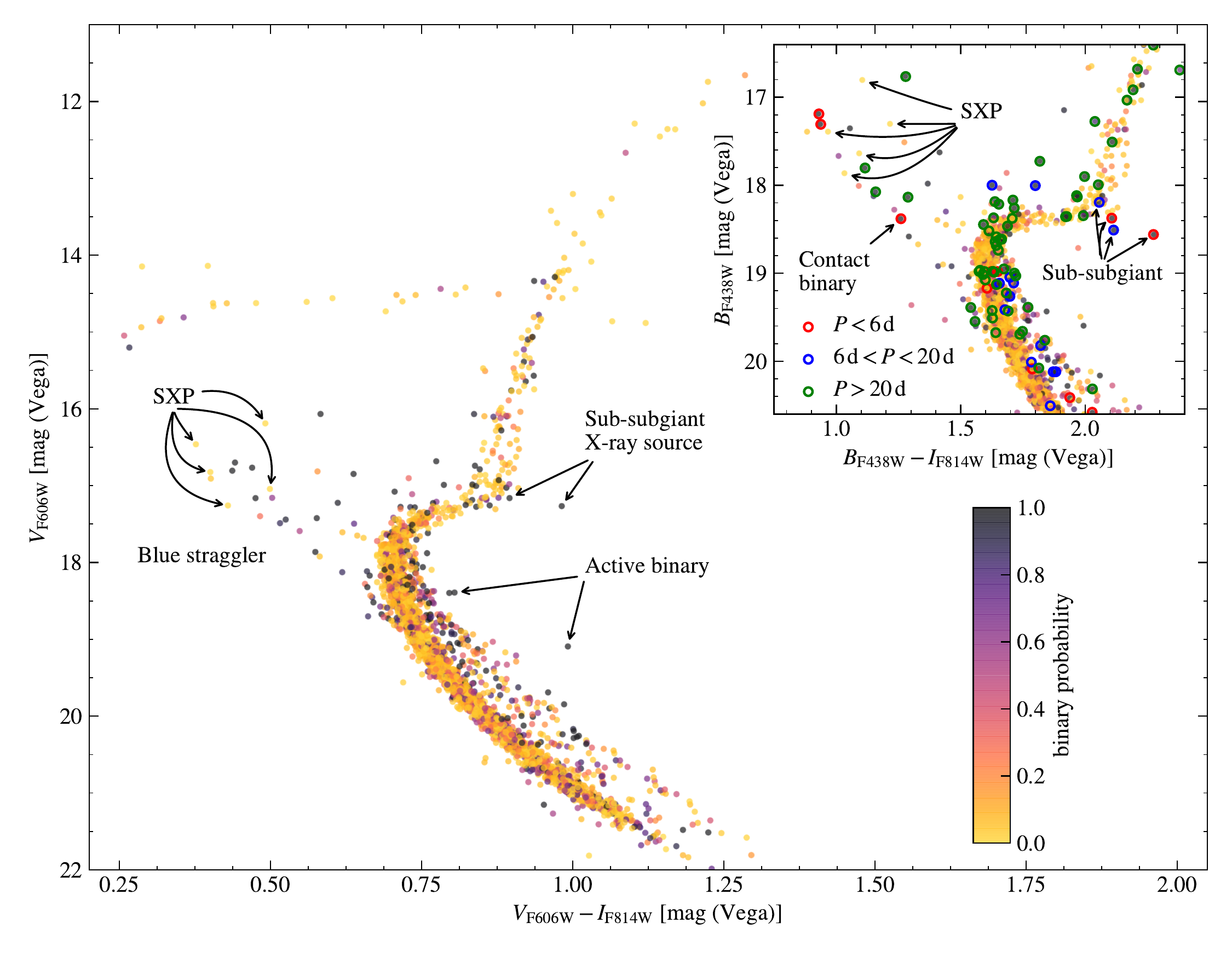}}
\caption{Colour-magnitude diagram of the members of \ngc{3201} created with the photometry taken from the HST UV \gc{} survey \citep{nardiello2018, piotto2015}. Colour-coded is the binary probability obtained by our statistical method. Large panel: the full CMD of our sample using V and I equivalent filters is shown. Small panel: a detailed version of the \ms{} turn-off CMD region using B and I equivalent filters is displayed. Additionally, the period $P$ range is indicated by coloured circles where \joker{} was able to fit a well constrained Keplerian orbit.}
\label{fig:cmd}
\end{figure*}

In \autoref{fig:joker_example} the result of this process for one example star (companion of a black hole candidate, more in \autoref{subsec:bh}) in our sample is presented. The 256 orbital solutions (in blue) represent the posterior likelihood distribution for this star and are in this case in principle two unique solutions with uncertainties. We define this solutions to be bimodal in orbital period. If in this example, only one solution (one cluster in the right panel of \autoref{fig:joker_example}) would be present, we would define it as unimodal. In general, we have unimodal samples when the standard deviation of \joker{} periods $P$ in logarithmic scale meets the criterion:
\begin{equation}
 \sigma_{\ln{P}} < 0.5 \;.
\label{eq:unimodal}
\end{equation}
To identify bimodal posterior samples we use a similar method as described in \citet{apogee}: We use the $k$-means clustering with $k = 2$ from the scikit-learn package \citep{scikit-learn} to separate two clusters in the posterior samples in orbital period (like the two clusters in the right panel of \autoref{fig:joker_example}). Each of the two clusters have to fulfill \autoref{eq:unimodal}. We take the most probable (the one with more samples than the other) as the result for this bimodal posterior samples. Finally, we take the median and the standard deviation as uncertainty of all parameters from these results.

%For 95 stars we were able to \changed{find a unimodal or bimodal Keplerian solution}.

\longtab{
\tiny
\begin{longtable}{llllS[table-format=4.4(5)]S[table-format=1.2(3)]S[table-format=3.1(3)]S[table-format=1.2(3)]S[table-format=1.2(3)]l}
\caption{\label{table:binaries} Binary system properties. The Keplerian parameters were calculated with \joker{}. The table is sorted by period.}\\
\hline\hline
{ACS Id} & {RA} & {Dec.} & {Mag.} & {Period} & {Eccentricity} & {Amplitude} & {Vis. mass} & {Invis. min. mass} & {Comment} \\
 & {\si{\degree}} & {\si{\degree}} & {F606W} & {\si{\day}} & & {\si{\kilo\meter\per\second}} & {\si{\solarmass}} & {$\si{\solarmass} \sin{i}$} &  \\
\hline
\endfirsthead
\caption{continued.}\\
\hline\hline
{ACS Id} & {RA} & {Dec.} & {Mag.} & {Period} & {Eccentricity} & {Amplitude} & {Vis. mass} & {Invis. min. mass} & {Comment} \\
 & {\si{\degree}} & {\si{\degree}} & {F606W} & {\si{\day}} & & {\si{\kilo\meter\per\second}} & {\si{\solarmass}} & {$\si{\solarmass} \sin{i}$} &  \\
\hline
\endhead
\hline
\endfoot
13108 & 154.40005 & -46.42014 & 17.68 & 0.269(7) & 0.50(21) & 33.4(143) & 1.20(5) & 0.11(5) & BSS, H$\alpha$, EW \\
25210 & 154.38502 & -46.40803 & 18.05 & 0.3731(1) & 0.22(24) & 9.2(23) & 0.80(5) & 0.03(1) & H$\alpha$, bimodal \\
12746 & 154.40330 & -46.41368 & 16.72 & 0.4706(1) & 0.01(2) & 28.0(8) & 1.20(5) & 0.12(1) & BSS, $V \sin{i}$ \\
13862 & 154.39460 & -46.42221 & 20.21 & 0.7287(1) & 0.09(12) & 41.9(64) & 0.60(5) & 0.14(3) & H$\alpha$, bimodal \\
22697 & 154.40615 & -46.41101 & 19.41 & 0.8631(1) & 0.12(10) & 30.7(24) & 0.69(5) & 0.12(2) & bimodal \\
12581 & 154.40447 & -46.41137 & 19.09 & 0.89(43) & 0.14(18) & 27.1(33) & 0.72(5) & 0.10(2) & bimodal \\
11222 & 154.41516 & -46.42245 & 16.82 & 0.9430(1) & 0.04(4) & 37.4(18) & 1.20(5) & 0.22(2) & BSS, $V \sin{i}$ \\
13362 & 154.39838 & -46.41245 & 20.86 & 0.9954(1) & 0.05(3) & 124.8(40) & 0.57(5) & 0.68(6) &  H$\alpha$ \\
12229 & 154.40736 & -46.41683 & 19.83 & 1.6900(3) & 0.10(14) & 15.0(29) & 0.66(5) & 0.07(2) & bimodal \\
10293 & 154.42364 & -46.41268 & 18.05 & 1.8760(1) & 0.04(3) & 66.3(17) & 0.80(5) & 0.44(3) &  \\
13968 & 154.39393 & -46.41370 & 19.29 & 1.9429(1) & 0.05(6) & 80.9(76) & 0.71(5) & 0.55(8) &  H$\alpha$ \\
22818 & 154.40547 & -46.39993 & 18.25 & 1.9477(2) & 0.24(24) & 14.7(57) & 0.79(5) & 0.08(3) & bimodal \\
11802 & 154.41090 & -46.41574 & 19.70 & 2.11(20) & 0.10(14) & 20.0(27) & 0.64(5) & 0.10(2) & H$\alpha$ \\
21859 & 154.41347 & -46.40691 & 20.37 & 2.2422(1) & 0.07(4) & 305.5(77) & 0.61(5) & 7.68(50) & BHC, H$\alpha$ \\
12836 & 154.40215 & -46.42134 & 20.13 & 2.4262(2) & 0.08(8) & 44.3(25) & 0.63(5) & 0.26(3) &  H$\alpha$ \\
22692 & 154.40662 & -46.39785 & 17.25 & 5.1038(4) & 0.02(3) & 38.6(16) & 0.82(5) & 0.35(3) & SSG, H$\alpha$+, X-ray \\
13438 & 154.39829 & -46.41403 & 17.17 & 5.9348(3) & 0.02(3) & 36.6(9) & 0.82(5) & 0.35(3) & SSG, H$\alpha$, X-ray \\
15138 & 154.38440 & -46.41622 & 18.14 & 6.393(2) & 0.05(9) & 18.8(23) & 0.79(5) & 0.16(3) & bimodal \\
11366 & 154.41408 & -46.41498 & 19.08 & 8.381(2) & 0.26(9) & 43.5(43) & 0.72(5) & 0.44(6) &  \\
23220 & 154.40234 & -46.41006 & 17.05 & 8.428(2) & 0.08(7) & 12.9(10) & 1.20(5) & 0.15(2) & BSS \\
25132 & 154.38563 & -46.41084 & 18.28 & 8.457(2) & 0.07(8) & 33.4(19) & 0.78(5) & 0.34(4) &  \\
10799 & 154.41927 & -46.41501 & 18.09 & 8.999(6) & 0.09(17) & 9.1(20) & 0.79(5) & 0.08(2) & bimodal \\
15222 & 154.38393 & -46.41161 & 18.77 & 9.963(2) & 0.26(9) & 22.8(19) & 0.75(5) & 0.22(3) &  \\
14749 & 154.38811 & -46.41873 & 17.03 & 10.006(2) & 0.09(7) & 42.9(15) & 0.82(5) & 0.53(4) & SSG, H$\alpha$, EA \\
10719 & 154.41981 & -46.42151 & 18.99 & 10.930(4) & 0.03(5) & 43.7(24) & 0.74(5) & 0.54(5) &  \\
12305 & 154.40665 & -46.41503 & 18.48 & 15.41(2) & 0.46(14) & 23.7(89) & 0.77(5) & 0.25(9) &  \\
23918 & 154.39586 & -46.40850 & 18.19 & 15.945(7) & 0.04(7) & 32.2(19) & 0.79(5) & 0.43(4) &  \\
10418 & 154.42208 & -46.42361 & 19.07 & 16.128(5) & 0.11(11) & 28.5(25) & 0.72(5) & 0.35(5) &  \\
13684 & 154.39605 & -46.42072 & 20.00 & 16.73(2) & 0.26(14) & 23.6(34) & 0.64(5) & 0.25(5) & bimodal \\
11405 & 154.41357 & -46.41997 & 17.25 & 17.219(6) & 0.42(8) & 14.1(14) & 0.82(5) & 0.15(2) & SSG, H$\alpha$ \\
10274 & 154.42387 & -46.41778 & 16.98 & 18.942(6) & 0.32(5) & 24.4(13) & 1.20(5) & 0.40(3) & BSS \\
12370 & 154.40612 & -46.41486 & 19.46 & 19.46(2) & 0.16(12) & 21.8(27) & 0.70(5) & 0.26(5) &  \\
23019 & 154.40364 & -46.40831 & 18.70 & 25.76(4) & 0.15(15) & 13.3(19) & 0.75(5) & 0.17(3) &  \\
23847 & 154.39665 & -46.40951 & 16.12 & 26.70(3) & 0.13(17) & 6.1(9) & 0.83(5) & 0.08(2) & bimodal \\
14837 & 154.38719 & -46.42433 & 19.12 & 32.37(5) & 0.43(18) & 16.0(20) & 0.71(5) & 0.20(4) &  \\
12706 & 154.40336 & -46.42295 & 18.05 & 34.11(4) & 0.27(10) & 12.8(13) & 0.80(5) & 0.18(3) &  \\
12871 & 154.40234 & -46.41310 & 20.35 & 38.2(2) & 0.09(14) & 18.3(27) & 0.61(5) & 0.27(11) & bimodal \\
12341 & 154.40620 & -46.42160 & 17.18 & 38.92(2) & 0.02(4) & 22.6(7) & 0.82(5) & 0.41(3) &  \\
21659 & 154.41506 & -46.41032 & 17.29 & 39.52(2) & 0.49(4) & 27.7(22) & 0.82(5) & 0.45(5) &  \\
13357 & 154.39872 & -46.41345 & 18.08 & 42.16(3) & 0.24(7) & 23.1(12) & 0.79(5) & 0.42(4) &  \\
12308 & 154.40673 & -46.41448 & 17.67 & 47.39(6) & 0.07(11) & 29.4(25) & 0.81(5) & 0.64(9) &  \\
12082 & 154.40831 & -46.42402 & 18.09 & 47.88(9) & 0.26(17) & 13.0(21) & 0.79(5) & 0.21(5) &  \\
22700 & 154.40637 & -46.41026 & 18.38 & 53.5(2) & 0.15(18) & 10.6(17) & 0.77(5) & 0.18(4) &  \\
13221 & 154.39961 & -46.41524 & 18.49 & 53.7(3) & 0.12(19) & 13.3(23) & 0.77(5) & 0.23(5) &  \\
13669 & 154.39623 & -46.42264 & 17.45 & 53.8(1) & 0.13(15) & 11.7(19) & 0.82(5) & 0.21(4) &  \\
24738 & 154.38877 & -46.41100 & 18.02 & 55.44(9) & 0.19(10) & 15.5(17) & 0.80(5) & 0.29(4) &  \\
25266 & 154.38429 & -46.40955 & 18.06 & 55.5(1) & 0.07(11) & 13.7(20) & 0.79(5) & 0.25(5) &  \\
13002 & 154.40101 & -46.41437 & 18.29 & 56.4(2) & 0.53(20) & 7.1(19) & 0.78(5) & 0.10(3) & bimodal \\
14889 & 154.38656 & -46.42310 & 18.67 & 56.71(6) & 0.39(15) & 27.8(85) & 0.76(5) & 0.56(19) &  \\
14972 & 154.38583 & -46.41625 & 18.02 & 69.2(2) & 0.18(13) & 12.2(17) & 0.80(5) & 0.23(5) &  \\
13756 & 154.39561 & -46.41871 & 18.52 & 71.3(3) & 0.72(16) & 32.3(200) & 0.77(5) & 0.53(33) &  \\
14743 & 154.38785 & -46.41935 & 18.20 & 74.3(4) & 0.36(12) & 20.5(24) & 0.79(5) & 0.43(8) &  \\
23000 & 154.40401 & -46.39918 & 17.61 & 75.3(3) & 0.20(21) & 6.3(12) & 0.81(5) & 0.12(3) &  \\
13522 & 154.39741 & -46.41123 & 18.09 & 108.2(5) & 0.28(9) & 17.7(23) & 0.80(5) & 0.44(8) &  \\
12904 & 154.40185 & -46.41983 & 16.07 & 110.3(3) & 0.04(6) & 11.6(7) & 1.20(5) & 0.35(3) & BSS, $V \sin{i}$ \\
14700 & 154.38856 & -46.41702 & 17.71 & 112.3(4) & 0.18(18) & 11.3(20) & 0.81(5) & 0.26(6) & bimodal \\
13355 & 154.39867 & -46.41374 & 19.93 & 115.1(6) & 0.63(12) & 34.9(100) & 0.67(5) & 0.82(22) &  \\
6324 & 154.38867 & -46.42641 & 15.64 & 120.5(3) & 0.57(12) & 14.2(48) & 0.83(5) & 0.29(8) &  \\
23175 & 154.40264 & -46.40494 & 15.07 & 129(2) & 0.12(20) & 5.2(14) & 0.83(5) & 0.12(4) &  \\
12927 & 154.40180 & -46.41472 & 19.07 & 143.4(6) & 0.37(9) & 14.1(15) & 0.72(5) & 0.33(5) &  \\
11300 & 154.41459 & -46.41824 & 17.00 & 163(2) & 0.41(12) & 17.3(44) & 0.82(5) & 0.48(13) &  \\
12560 & 154.40454 & -46.41537 & 17.57 & 167.01(9) & 0.61(2) & 72.3(17) & 0.81(5) & 4.53(21) & BH1 \\
4513 & 154.40662 & -46.42597 & 17.68 & 172(2) & 0.07(12) & 11.4(16) & 0.81(5) & 0.32(6) &  \\
22751 & 154.40602 & -46.40024 & 14.35 & 172.6(6) & 0.03(4) & 10.5(7) & 0.83(5) & 0.30(3) &  \\
12658 & 154.40360 & -46.42207 & 15.40 & 176.0(4) & 0.07(5) & 13.4(7) & 0.83(5) & 0.40(4) &  \\
14355 & 154.39107 & -46.41961 & 17.15 & 191(3) & 0.06(15) & 11.1(15) & 1.20(5) & 0.41(7) & BSS, bimodal \\
12438 & 154.40560 & -46.41283 & 17.22 & 197(2) & 0.06(10) & 8.2(15) & 0.82(5) & 0.23(5) &  \\
23276 & 154.40164 & -46.40921 & 17.51 & 204.2(10) & 0.89(6) & 21.2(106) & 0.82(5) & 0.30(12) &  \\
23452 & 154.40018 & -46.40326 & 16.79 & 206(2) & 0.06(8) & 7.2(9) & 0.82(5) & 0.20(3) &  \\
12975 & 154.40134 & -46.42046 & 18.67 & 224(4) & 0.07(11) & 11.5(15) & 0.77(5) & 0.35(6) &  \\
13391 & 154.39790 & -46.42029 & 18.71 & 225(6) & 0.24(25) & 9.8(22) & 0.76(5) & 0.28(8) &  \\
23381 & 154.40051 & -46.40725 & 17.24 & 262(2) & 0.11(15) & 12.1(17) & 0.82(5) & 0.42(7) &  \\
23889 & 154.39636 & -46.40059 & 17.64 & 282(5) & 0.33(14) & 13.2(20) & 0.81(5) & 0.43(9) &  \\
11945 & 154.40948 & -46.41194 & 17.44 & 284(8) & 0.18(20) & 9.8(27) & 1.20(5) & 0.40(15) & BSS \\
14055 & 154.39350 & -46.41191 & 17.27 & 286(5) & 0.17(18) & 9.0(16) & 0.82(5) & 0.29(6) &  \\
13816 & 154.39521 & -46.41600 & 16.11 & 302(5) & 0.17(20) & 3.9(10) & 0.83(5) & 0.11(3) &  \\
23848 & 154.39645 & -46.40946 & 16.52 & 323(2) & 0.48(8) & 15.6(15) & 1.20(5) & 0.67(11) & BSS \\
12173 & 154.40785 & -46.41662 & 17.77 & 340(6) & 0.47(24) & 20.0(79) & 0.81(5) & 0.80(30) & bimodal \\
13069 & 154.40069 & -46.41497 & 17.01 & 375(3) & 0.14(19) & 9.3(27) & 0.82(5) & 0.35(14) &  \\
12932 & 154.40168 & -46.41370 & 17.81 & 386(9) & 0.48(20) & 6.5(19) & 0.81(5) & 0.19(7) &  \\
11831 & 154.41014 & -46.42258 & 13.88 & 493(11) & 0.07(14) & 6.4(9) & 0.83(5) & 0.25(4) &  \\
24135 & 154.39382 & -46.41077 & 18.12 & 509(20) & 0.51(30) & 8.2(67) & 0.79(5) & 0.28(18) &  \\
13587 & 154.39637 & -46.42422 & 18.52 & 537(21) & 0.13(15) & 10.6(18) & 1.20(5) & 0.59(14) & BSS \\
11317 & 154.41447 & -46.41358 & 15.77 & 602(12) & 0.25(13) & 9.4(14) & 0.83(5) & 0.42(7) &  \\
13874 & 154.39476 & -46.42003 & 18.59 & 608(27) & 0.30(30) & 10.1(40) & 0.77(5) & 0.43(20) &  \\
11779 & 154.41062 & -46.42064 & 17.43 & 620(35) & 0.13(21) & 8.2(17) & 0.82(5) & 0.36(10) &  \\
15293 & 154.38311 & -46.42006 & 16.75 & 668(20) & 0.48(10) & 11.3(18) & 0.82(5) & 0.49(10) &  \\
5132 & 154.40034 & -46.42585 & 20.17 & 764(11) & 0.28(16) & 36.3(84) & 0.64(5) & 4.40(282) & BHC, bimodal \\
13782 & 154.39552 & -46.41252 & 17.26 & 846(61) & 0.11(20) & 8.2(19) & 0.82(5) & 0.42(12) &  \\
25211 & 154.38487 & -46.40768 & 17.41 & 988(75) & 0.11(16) & 8.7(17) & 1.20(5) & 0.59(14) & BSS \\
12828 & 154.40192 & -46.42344 & 18.09 & 1075(37) & 0.05(8) & 13.2(14) & 0.80(5) & 0.90(16) &  \\
11131 & 154.41636 & -46.41623 & 17.28 & 1627(222) & 0.31(17) & 8.2(16) & 0.82(5) & 0.54(13) & bimodal \\
23215 & 154.40267 & -46.39830 & 18.73 & 1997(304) & 0.08(14) & 16.3(21) & 0.76(5) & 1.82(39) & NSC \\
13154 & 154.39996 & -46.41225 & 18.17 & 2773(1009) & 0.16(19) & 10.1(41) & 0.79(5) & 0.94(94) &  \\
13808 & 154.39505 & -46.41923 & 15.23 & 3001(1005) & 0.11(22) & 3.2(14) & 0.83(5) & 0.21(13) &  \\
\end{longtable}
\tablefoot{
\textbf{ACS Id}: Identifier in the catalogue of \ngc{3201} in the ACS \gc{} survey \citep{acs1}.
\textbf{BH1}: Published black hole in \citet{giesers}.
\textbf{BHC}: Star with a companion that could be a black hole (see \autoref{subsec:bh}).
\textbf{BSS}: \bs{} star (see \autoref{subsec:bss_sxp}).
\textbf{bimodal}: The posterior sampling of the periods found by \joker{} shows two modes.
\textbf{EA}: Detached eclipsing binary \citep{clement2017}.
\textbf{EW}: Contact eclipsing binary \citep{kaluzny2016}.
\textbf{H$\alpha$}: Compared to our best fitting model spectrum (and to a well fitted H$\beta$ line) several spectra of this star show a partially filled-in H$\alpha$ absorption line (absorption is not as deep as expected).
\textbf{H$\alpha$+}: Star showing real H$\alpha$ emission in several MUSE spectra.
\textbf{NSC}: Star with a companion that could be a neutron star.
\textbf{SSG}: Sub-subgiant star (see \autoref{subsec:ssg}).
$\pmb{V \sin{i}}$: star with measured spin velocity according to \citet{simunovic2014} (see \autoref{subsec:bss_sxp}).
\textbf{X-ray}: Star listed as X-ray source in the November 2017 pre-release of the Chandra Source Catalog Release 2.0 \citep{evans2010}.
}
}

\subsection{Binary system properties}
\label{sec:binaries}
The CMD in \autoref{fig:cmd} is created using the newer HST UV \gc{} survey photometry of \ngc{3201} \citep{nardiello2018, piotto2015} which was matched to the photometry of the HST ACS \gc{} survey \citep{acs1,acs2} we used as an input catalogue for the extraction of the spectra. The binary probability for each star we determined by our statistical method is colour-coded. As expected, many \ms{} binaries positioned to the red of the \ms{} are found by our method. Red giants brighter than the horizontal branch magnitude could not be found in binary systems in our sample. A good check for the statistical method are the chromospherically active binaries, which deviate in our CMD significantly from the \ms{} (see annotation in CMD) and are very well confirmed by the method. More details about the other stellar types are presented in the following \autoref{sec:discussion}.

As described in \autoref{sec:joker} we analysed all stars with a \SI{50}{\percent} variability probability. We found 95 stars from which 78 stars have unimodal and 17 stars bimodal posterior samples in orbital period. That means the period of these stars is well constrained but does not mean all Keplerian parameters are constrained similarly well. We present the results of \joker{} in \autoref{table:binaries} with the star position, magnitude, a selection of the fitted parameters (period, eccentricity, amplitude, visible mass), and the derived invisible minimum (companion) mass. The comment column contains additional information to individual stars as explained in the table notes. For a selection of interesting stars of this table we show the best Keplerian fit in \autoref{fig:joker_interesting} and a blind random set in \autoref{fig:joker_random}. These plots show in the upper panel the radial velocities $v_r$ of an individual star from our final sample phase folded with the period from the best-fitting model. The lower panel contains the residuals after subtracting this model from the data. In addition to the identifier (ACS Id), position and magnitude from the ACS catalogue \citep{acs1}, we note the period $P$, eccentricity $e$, invisible minimum mass $M_2$, \textit{reduced} $\chi^2$ of the best fitting model, and the comment in every plot for convenience.

\begin{figure}
\resizebox{\hsize}{!}{\includegraphics{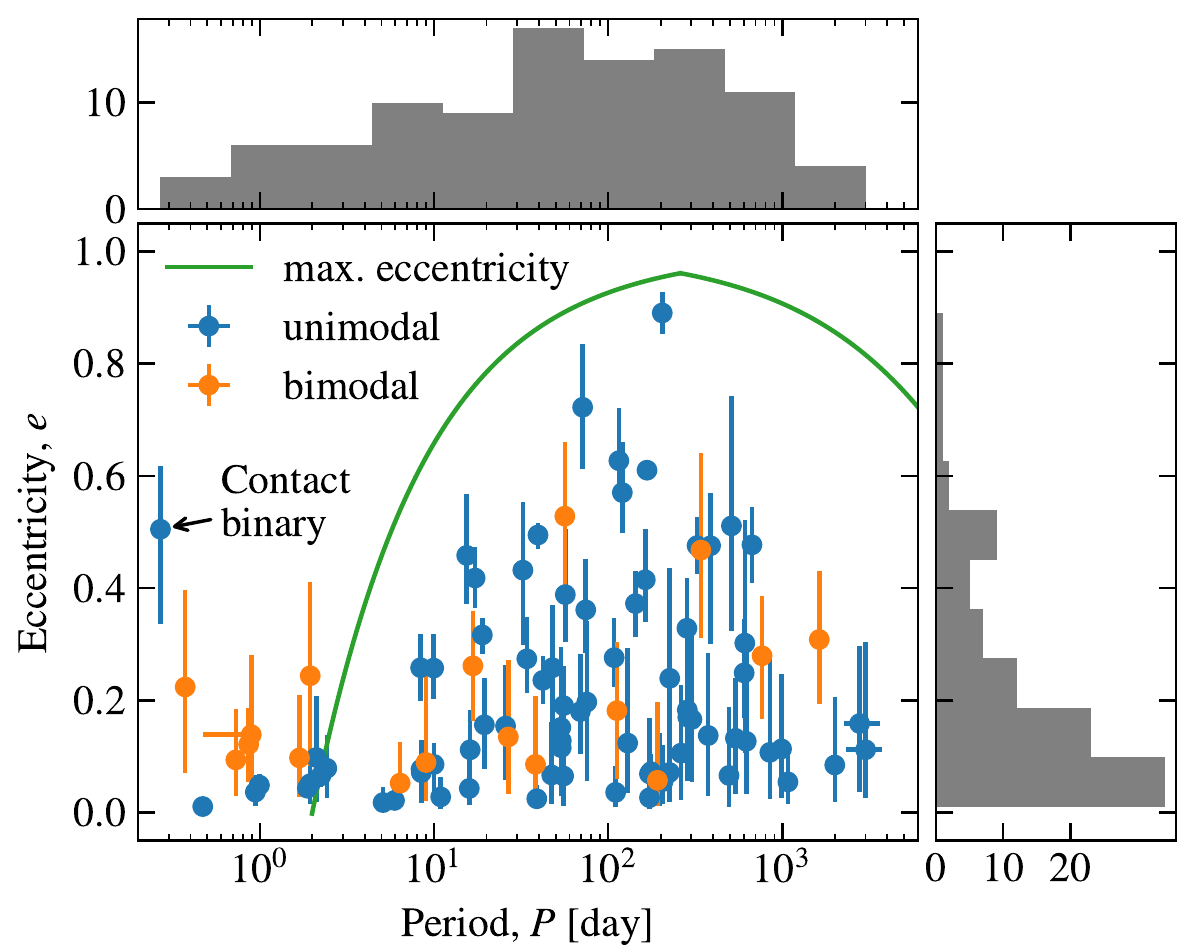}}
\caption{Eccentricity-period plot of the well constrained binaries in \ngc{3201}. Binaries with unimodal (blue) and bimodal (orange) solutions in the posterior period sampling. The maximum eccentricity in green represents the theoretical limit (above a \SI{2}{\day} period) to which binary systems are stable (see \autoref{sec:binaries}).}
\label{fig:binaries_ecc_vs_period}
\end{figure}

\begin{figure}
\resizebox{\hsize}{!}{\includegraphics{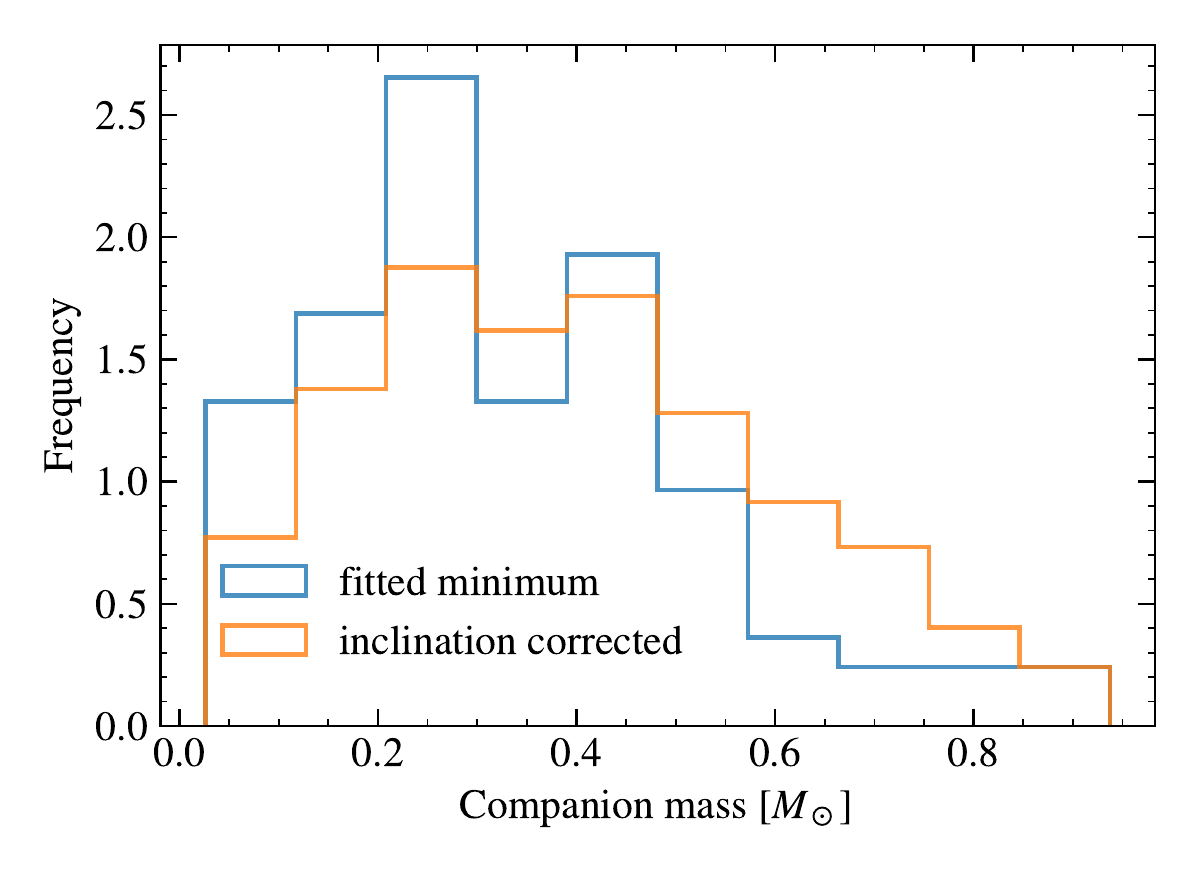}}
\caption{The companion masses (< \SI{1}{\solarmass}) of the well constrained binaries in \ngc{3201} are shown. The blue histogram contains the fitted results (minimum masses) from \joker{}, whereas the orange histogram contains the same results after applying a statistical correction for orbital inclination to obtain the actual companion mass distribution (see \autoref{sec:binaries}).}
\label{fig:binaries_hist_companion}
\end{figure}

For the first time \autoref{fig:binaries_ecc_vs_period} shows the eccentricity-period distribution of binaries in a \gc{}. We found binaries in a period range from \SIrange{0.27}{3000}{\day} with eccentricities from 0~to~0.9. The eccentricity distribution is biased towards low eccentricities, because for these orbits fewer measurements are necessary to get a unique solution. In the figure we also show a maximum eccentricity $e_\textrm{max}$ power law derived from a Maxwellian thermal eccentricity distribution
\begin{equation}
 e_\textrm{max}(P) = 1 - \left(\frac{P}{2\;\textrm{days}}\right)^{-2/3} \quad \textrm{for}\;P>2\;\textrm{days} 
\end{equation}
for a given period $P$ \citep{moe2017}. This represents the binary components having Roche lobe fill factors $\leq \SI{70}{\percent}$ at periastron. All binaries with $P < \SI{2}{\day}$ should have circular orbits due to tidal forces. Additionally, we also take dynamical star interactions into account by using a limit on the orbital velocity at apoastron. The orbital velocity should always be higher than the central cluster dispersion of \SI{5}{\kilo\meter\per\second} of \ngc{3201} \citep[][2010 edition]{harris}. This limit on eccentricity is dominant for periods $> \SI{260}{\day}$. Except for the contact binary star (ACS Id: \#13108, see \autoref{table:binaries}) all stars behave accordingly. Note that when the eccentricity distribution of one star peaked at zero we still took the median like in all other distributions. Therefore the stars below $P < 2$ are consistent within their uncertainty with $e = 0$. \Autoref{fig:binaries_ecc_vs_period} also clearly shows that not all binaries in \gc{}s have been circularised over the lifetime of a \gc{}. We, however, do not find high-eccentric long-period (e.g. $e > 0.6$ and $P > \SI{300}{\day}$) binaries within the eccentricity limit. This could either be due to our observational bias or an effect which restricts the eccentricity even further. %For example, a binary with $P = \SI{1000}{\day}$ and $e = \num{0.8}$ would have a separation of $\SI{3.85}{\au}$ at apoastron which is within the influence sphere of other stars during the lifetime of \ngc{3201} \citep{hut1992}.

\Autoref{fig:binaries_hist_companion} shows the minimum companion masses of the well constrained binaries (blue histogram). In orange all masses are resampled 1000 times from their fitted mass divided by $\sin{i}$ to obtain the actual companion mass distribution. The inclination $i$ is taken from a sinusoidal detection probability distribution between 0 to $\pi/2$ with a cutoff mass of \SI{0.8}{\solarmass} (roughly maximum normal stellar mass in \ngc{3201}). This results in a linear anti-correlation between the companion mass and the frequency of binaries having this companion mass. As explained in \autoref{sec:simulation} we have a bias towards binaries with both components having different luminosities. \citet{milone2012} found a flat mass ratio distribution throughout all clusters above a mass ratio $> 0.5$. The pairing fraction of our binaries is beyond the scope of this paper, since we would need to correct for the luminosity ratio effect and the selection function of \joker{}, which is currently unknown in our case.

Noteworthy is that all red giant binaries in our sample with well constrained orbits have periods larger than \SI{100}{\day}. This corresponds to a minimum semi-major axis of $\gtrsim \SI{0.5}{\au}$ and is consistent with not having Roche-lobe overflow in the binary system.

%Our final MUSE sample for the binary analysis consists of \num{3553} stars (see \autoref{sec:finalsample}). From these, \changed{following a simple assignment in the CMD}, \num{3229} are \ms{} stars, \num{130} subgiants, \num{134} red giants, \num{20} horizontal branch stars, and \num{40} \bss{}s. For this full final sample we obtained a binary probability and for 95 stars we were able to fit a Keplerian orbit to the data.

\section{Peculiar objects in \ngc{3201}}
\label{sec:discussion}

\subsection{\Bss{}s and \SXP{}-type stars}
\label{subsec:bss_sxp}
Our sample of \ngc{3201} contains 40 \bs{}s with sufficient observations and signal to noise. From these stars, 23 are likely in binary systems while 17 are not (including the \SXP{}-type stars assumed to be single stars). This observational binary fraction of \bbf{} is much higher than the one of all cluster stars or even the Milky Way field stars (see \autoref{sec:introduction}). Our ratio of binaries to single \bss{}s $R_\textrm{B/S} = 1.35$ in the core of \ngc{3201} is significantly higher than the prediction by \citet{hypki2016} of $R_\textrm{B/S} \sim 0.4$, but could also be explained by an overabundance of the more massive \bs{} binaries (compared to single \bs{}s) due to mass segregation. Nevertheless, this high binary fraction strongly suggests that most of the \bs{}s in our sample were formed within a binary system. As in \autoref{sec:peculiar} explained, a cluster member needs additional mass to become a \bs{}. Two formation scenarios are consistent with such a high binary fraction: (1) mass transfer within a triple star system \citep[e.g.][]{antonini2016}, (2) stellar mergers induced by stellar interactions of binary systems with other binaries or single stars \citep[e.g.][]{leonard1989,fregeau2004}.

From these 23 \bs{} binaries \joker{} could find 11 highly constrained solutions, three are extremely hard binaries (< \SI{1}{\day}) with a minimum companion mass ranging from \SIrange{0.11}{0.24}{\solarmass} and six relatively wide (> \SI{100}{\day}) with minimum companion masses ranging from \SIrange{0.35}{0.67}{\solarmass}. The remaining 2 \bs{}s are in between the period and companion mass range (see stars in \autoref{table:binaries} with comment 'BSS' for more details). \Autoref{fig:cmd} shows the \bs{}s and their period range in a colour-magnitude diagram. We also identified five \bss{}s as known \SXP{}-type variable stars in the \textit{Catalogue of Variable Stars in Galactic Globular Clusters} \citep{clement2001, clement2017} and one of them as a \SXP{} candidate in \citet{arellano2014}. We also see their pulsations (similar to the \RRL{}-type stars) as a radial velocity signal up to amplitudes of \SI{18}{\kilo\meter\per\second}. We cannot distinguish between the pulsation signal and a possible signal induced by a binary companion, thus we consider these stars to be single stars and the velocity variability solely explained by radial pulsations.

Mass transfer in a relatively wide \bs{} binary system is almost impossible. Thus, only the three hard binaries are likely to result from this process. It is obvious that mass transfer is taking place in the contact binary system with ACS Id \#13108. On the other hand, head-on collisions of single stars would result in single \bs{}s and are extremely unlikely in a low-density cluster such as \ngc{3201}. Therefore, it is more likely that a significant fraction of the \bs{}s in our sample result from coalescence in systems with $\geq3$ stars. The complex behaviour of such systems favours configurations that result in the coalescence of two of its members \citep[e.g.][]{leonard1989,fregeau2004,antonini2016}. If more than two stars in the formation of \bs{}s are involved, this could also mean that hard \bs{} binary systems could have a third outer component inducing an additional, much weaker signal. We could not identify such a signal in our data. For example, the star with ACS Id \#12746 (see radial velocity signal in \autoref{fig:joker_interesting}) is the most central hard \bs{} binary in our sample with 63 observations, but its radial velocity curve does not contain any significant additional signal.

%A mass transfer in a relatively wide \bs{} binary system is almost impossible. \changed{Thus, the three wide \bs{}s in our sample could only be explained by a stellar merger, since the inner component is not present anymore. In the case of the triple star formation scenario, these \bs{}s} were triple star systems in the first place and the inner (mass-transfering) components have merged. This triple star formation mechanism of \bs{}s could be more efficient through the perturbations by the third outer star compared to a (tidal) capture in a binary system \citep[see][]{antonini2016}. \changed{In the case of the binary-binary interaction formation scenario, temporarily a chaotic quadruple system could be formed in which two components merge, one component is taking the angular momentum away from the system and one component could survive to form a new binary system with the merged \bs{}.

The projected spin velocities ($V \sin{i}$) of \bs{}s have been suggested as a possibility to infer their ages and origins \citep[e.g.][]{leiner2018}. $V \sin{i}$ measurements of \bs{}s in \ngc{3201} have been performed by \citet{simunovic2014} and cross-matching their sample with our data results in a subset of five \bs{}s with spin velocities ranging from \SI{34.6(16)}{\kilo\meter\per\second} to \SI{135.2(35)}{\kilo\meter\per\second}. We note that all of them are binary systems. However, no clear picture emerges regarding a possible relation between spin velocity and orbital properties.

As introduced, \SXP{}-type stars are likely formed in binary evolution. That is why it is possible for them to have a companion, but the companions radial velocity signal would be hidden in the pulsation signal.

\subsection{\Ssg{} stars}
\label{subsec:ssg}
We found four \ssg{} (SSG) stars\footnote{The four \ssg{}s in our sample could also be defined as red stragglers, since they are at the boundary of both definitions in \citet{geller2017a}.} within our MUSE FoV (see position in \autoref{fig:cmd}). Assuming the radial velocity changes are only Keplerian (and not by pulsations), \joker{} could determine unique Keplerian solutions for all of them (see stars in \autoref{table:binaries} with comment 'SSG').

Fortunately the SSG with ACS Id \#14749 (see also plot in \autoref{fig:joker_interesting}) is a known detached eclipsing binary with a period of \SI{10.0037}{\day} \citep{kaluzny2016} and we found the same period \SI{10.006(2)}{\day} with our method. But this also means the system is observed edge-on and our companion mass is its true mass. The visible star shows a radial velocity amplitude of $\SI{42.9(15)}{\kilo\meter\per\second}$ with a very low eccentricity of $\num{0.09(7)}$, and a companion mass of $\SI{0.53(4)}{\solarmass}$ for an assumed primary mass of $\SI{0.82(5)}{\solarmass}$. That makes a system mass of \SI{1.35}{\solarmass} and considering mass transfer, one possible pathway could be towards the \bs{} region \citep{leiner2017}. Compared to our best fitting model spectrum (and to a well fitted H$\beta$ line) all spectra of this star show a partially filled-in H$\alpha$ absorption line (the absorption is not as deep as expected, see \autoref{fig:ssg_filled} as an example). This could mean that mass transfer is already underway. In this case this could be used to infer an accretion rate, but is beyond the scope of this paper.

The two short period ($ < \SI{6}{\day}$, \#22692, \#13438) \ssg{}s show X-ray emission according to the November 2017 pre-release of the Chandra Source Catalog Release 2.0 \citep{evans2010}. The SSG (\#22692) with the \SI{5.1}{\day} period actually has significant varying H$\alpha$ emission lines in most of our spectra. One spectrum of this star with H$\alpha$ in emission is presented in \autoref{fig:ssg_emission}. The maximum emission line is twice as strong as the minimum line compared to the continuum. Again, it would be interesting to calculate the accretion rate for this case, but is beyond the scope of this paper. The SSG (\#13438) with the \SI{5.9}{\day} period has partially filled-in H$\alpha$ absorption lines in all spectra like the eclipsing SSG described before. One spectrum of this star with the filled-in H$\alpha$ line is shown in \autoref{fig:ssg_filled}. Both short period \ssg{}s have a similar semi-amplitude of $\sim \SI{37}{\kilo\meter\per\second}$, no eccentricity and a minimum companion mass of \SI{0.35(3)}{\solarmass}.

The SSG (\#11405) has with \SI{17.2}{\day} the longest period, unlike the other stars a relatively high eccentricity of 0.4 and a small minimum companion mass of \SI{0.15}{\solarmass}.

The detection of a dozen more SSGs and RSs in our MUSE \gc{} sample will be published in the emission line catalogue of G\"ottgens et al. subm.

\subsection{Black hole candidates}
\label{subsec:bh}
%\begin{figure}
%\resizebox{\hsize}{!}{\includegraphics{figures/1406607}}
%\caption{The top panel shows the radial velocity measurements $v_\mathrm{r}$ of the black hole candidate, phase folded with the period. Green points are additional measurements since the publication of \citet{giesers}. Error bars are smaller than the data points. The red curve shows the already published best-fitting Keplerian orbit. The middle panel contains the residuals after subtracting this best-fitting model to the data. The bottom panel shows the radial velocity measurements of the reference star. Grey dots are phase shifted duplicates of the black ones, and are included to improve the visualization.}
%\label{fig:bh}
%\end{figure}

Here we report our new measurements for the stellar-mass black hole (BH) candidate in \ngc{3201} previously published in \citet{giesers}. The additional observations perfectly match the known orbital model and confirm the BH minimum mass to be \bh{}. The deviation to the previously minimum mass of \SI{4.36(41)}{\solarmass} is within the uncertainties. Furthermore, the observations fill in the apoapsis which was missed by previous observations (see \#12560 in \autoref{fig:joker_interesting}) and exclude other possible orbital solutions. 
We would like to emphasise that this star has been analysed with \joker{} and MCMC like all other stars, without the effort invested in it as in \citet{giesers}.

In our full sample of \ngc{3201} we detected two additional BH candidates with solutions from the fit of \joker{}. \Autoref{table:binaries} shows the resulting properties of these binary systems with ACS Id \#21859 and \#5139. Both systems are faint with extracted spectra of S/N~$\leq 10$. Remarkable is the system (\#21859) with a unique solution and a semi-amplitude of about \SI{300}{\kilo\meter\per\second}, we cannot explain the radial velocity curve shown in \autoref{fig:joker_interesting} with other explanations than an unseen companion with a minimum mass of \SI{7.68(50)}{\solarmass}. The other system (\#5132) has two solutions with the more probable (\SI{82}{\percent} probability) minimum companion mass of \SI{4.40(282)}{\solarmass} and the less probable of \SI{1.10(20)}{\solarmass}. This candidate is not well constrained and needs more observations to be confirmed as a BH. To our knowledge, both sources do not have an X-ray or radio counterpart. This is remarkable in the case of \#21859, as the presence of a partially filled-in H$\alpha$ line in the MUSE spectra suggests that accretion is present. We show a combination of all spectra shifted to rest-frame in \autoref{fig:bh_candidate}.

In light of the recent results based on Monte Carlo simulations, it appears feasible that \ngc{3201} hosts several BHs in binaries with \ms{} stars. Both \citet{kremer2018} and \citet{askar2018b} predict the presence of at least $\sim100$ BHs in the cluster, with tens of them in binaries with other BHs or bright companions. More specifically, in the best model of \citet{kremer2018}, four BHs have \ms{}-star companions. Interestingly, all of those systems show eccentricities $\gtrsim0.6$, add odds with the new candidates that we discovered (cf. \autoref{table:binaries}). On the other hand, the semi-major axes of our candidates of \SI{0.067}{\au} and \SI{2.804}{\au} are within the range of semi-major-axis values that \citet{kremer2018} found in their best model of \ngc{3201} (see their Fig.~3).

Our \mocca{} snapshot at \SI{12}{Gyr} contains 43 BHs with 33 single BHs, 2 BH-BH binaries, and 6 BH-star binaries. The median mass of all BHs is \SI{16.8}{\solarmass} within a range of \SI{3.7}{\solarmass} to \SI{78}{\solarmass}. Two BH-star binaries and 5 single BHs are in the probable mass region of the BH candidates we discussed here. The total number of BHs is lower than the previous predictions, but the number of BHs with detectable companions is consistent with our observations throughout all models.

\section{Conclusions and outlook}
\label{sec:conclusions}
We elaborated the first binary study from our MUSE survey of 27 Galactic \gc{}s on the core of \ngc{3201} and showed the variety of results our blind multi-epoch spectroscopic observations reveal. Since modelling a \gc{} contains many imponderables, we developed a statistical method -- applicable to any time-variant measurements with inhomogeneous sampling and uncertainties (not only radial velocities) -- which can distinguish between a constant and varying signal (like in single and binary stars). We determined an observational binary frequency of \obf{} using this statistical method. Based on a comparison with an advanced \mocca{} simulation of \ngc{3201}, we calculated a total binary fraction of \cbf{} for all stars in the whole cluster. We confirmed the trend of an increasing binary fraction towards the cluster centre due to mass segregation. We found a significantly higher binary fraction \bbf{} of \bss{}s compared to the cluster, indicating that the formation of \bs{}s in the core of \ngc{3201} is related to binary evolution. For the first time, we presented well constrained Keplerian orbit solutions for a significant amount of stars (95) using the Monte Carlo tool \joker{}. Eleven of these are \bss{}s likely in binary systems and shed light on the properties of these systems. We conclude that both the mass transfer formation scenario and the collisional formation scenario of \bs{}s are present in our data. Collision means in this case the coalescence of two stars during a binary-binary or binary-single encounter. We found four \ssg{} stars by connecting our MUSE spectroscopy with HST photometry and X-ray observations. Fortunately, we got definite Keplerian solutions for all of them and have insights into their properties for the first time in a \gc{}. Finally, we presented three stellar-mass black hole candidates, from which one is already published \citep{giesers} and one with a minimum mass of \SI{7.68(50)}{\solarmass} is clearly above a single and even binary neutron star companion mass limit. In total, these black hole candidates in binary systems with \ms{} stars would strongly support the hypothesis that \ngc{3201} has an extensive black hole population of up to hundred more black holes \citep{kremer2018,askar2018b}. This cluster, and maybe other \gc{}s as well, could be a significant source of gravitational waves.

In a following paper we will present binary fractions in the context of multiple stellar populations within \ngc{3201}. We will continue to observe $\omega$~Cen and 47~Tuc to have a comparable amount of epochs and will end up with a magnitude more stars per cluster compared to \ngc{3201}. Finally, we will publish the binary fractions of all clusters in our survey and try to find correlations with cluster parameters.

\begin{acknowledgements}
We thank Arash Bahramian, Nate Bastian, Robert Mathieu, Wolfram Kollatschny, Stan Lai, and Mark Gieles for helpful discussions. BG, SD, SK and PMW acknowledge support from the German Ministry for
Education and Science (BMBF Verbundforschung) through grants 05A14MGA,
05A17MGA, 05A14BAC, and 05A17BAA. AA is supported by the Carl Tryggers Foundation for Scientific Research through the grant CTS 17:113. SK gratefully acknowledges funding from a European Research Council consolidator grant (ERC-CoG-646928- Multi-Pop). JB acknowledges support by FCT/MCTES through national funds by grant UID/FIS/04434/2019 and through Investigador FCT Contract No. IF/01654/2014/CP1215/CT0003. This research is supported by the German Research
Foundation (DFG) with grants DR 281/35-1 and KA 4537/2-1. Based on observations made with ESO Telescopes at the La Silla Paranal Observatory under programme IDs 094.D-0142, 095.D-0629, 096.D-0175, 097.D-0295, 098.D-0148, 0100.D-0161, 0101.D-0268, 0102.D-0270, and 0103.D-0204.
Based on observations made with the
NASA/ESA Hubble Space Telescope, obtained from the data archive at the
Space Telescope Science Institute. STScI is operated by the
Association of Universities for Research in Astronomy, Inc. under NASA
contract NAS 5-26555.
Supporting data for this article is available at: \url{http://musegc.uni-goettingen.de}.
\end{acknowledgements}

%% file: 3201binaries_appendix.tex
\begin{table*}
\caption{Radial velocity measurements: 10 randomly selected rows from the electronically published catalogue with $\sim \num{50000}$ rows.}
\label{table:catalogue}
\centering
\begin{tabular}{c c c c c c c c}
\hline\hline
ACS Id & RA & Dec. & Mag. & BMJD & $v_\textrm{r}$ & $\epsilon_v$ & $P(\chi_i^2, \nu_i)$ \\
 & {\si{\degree}} & {\si{\degree}} & {F606W} & {\si{\day}} & {\si{\kilo\meter\per\second}} & {\si{\kilo\meter\per\second}} & {\%}\\
\hline
11317 & 154.41447 & -46.41358 & 15.77 & 2457136.4895 & 491.7 & 1.5 & 100 \\
24879 & 154.38805 & -46.40288 & 19.86 & 2457009.7918 & 499.4 & 6.4 & 36 \\
10605 & 154.42077 & -46.42044 & 18.52 & 2458490.7212 & 481.6 & 3.3 & 17 \\
12088 & 154.40818 & -46.42113 & 19.09 & 2457009.8027 & 491.0 & 4.7 &  4 \\
15271 & 154.38357 & -46.41156 & 20.50 & 2457786.8727 & 496.3 & 7.2 & 16 \\
23145 & 154.40307 & -46.39943 & 20.74 & 2457008.8456 & 490.3 & 7.2 & 60 \\
11426 & 154.41367 & -46.41481 & 18.17 & 2458227.6446 & 506.6 & 8.0 & 35 \\
22284 & 154.41001 & -46.39965 & 20.39 & 2458227.6446 & 517.2 & 8.3 & 13 \\
22723 & 154.40600 & -46.40560 & 18.70 & 2458490.7325 & 478.0 & 4.8 &  7 \\
13914 & 154.39437 & -46.41315 & 21.16 & 2458249.5006 & 495.6 & 8.1 &  7 \\
\hline
\end{tabular}
\tablefoot{
\textbf{ACS Id}: Identifier in the catalogue of \ngc{3201} in the ACS \gc{} survey \citep{acs1}. Position and magnitude are also taken from this catalogue.}
\end{table*}

\begin{figure*}
\centering{\textbf{Selection of Joker results}}
\resizebox{\hsize}{!}{\includegraphics{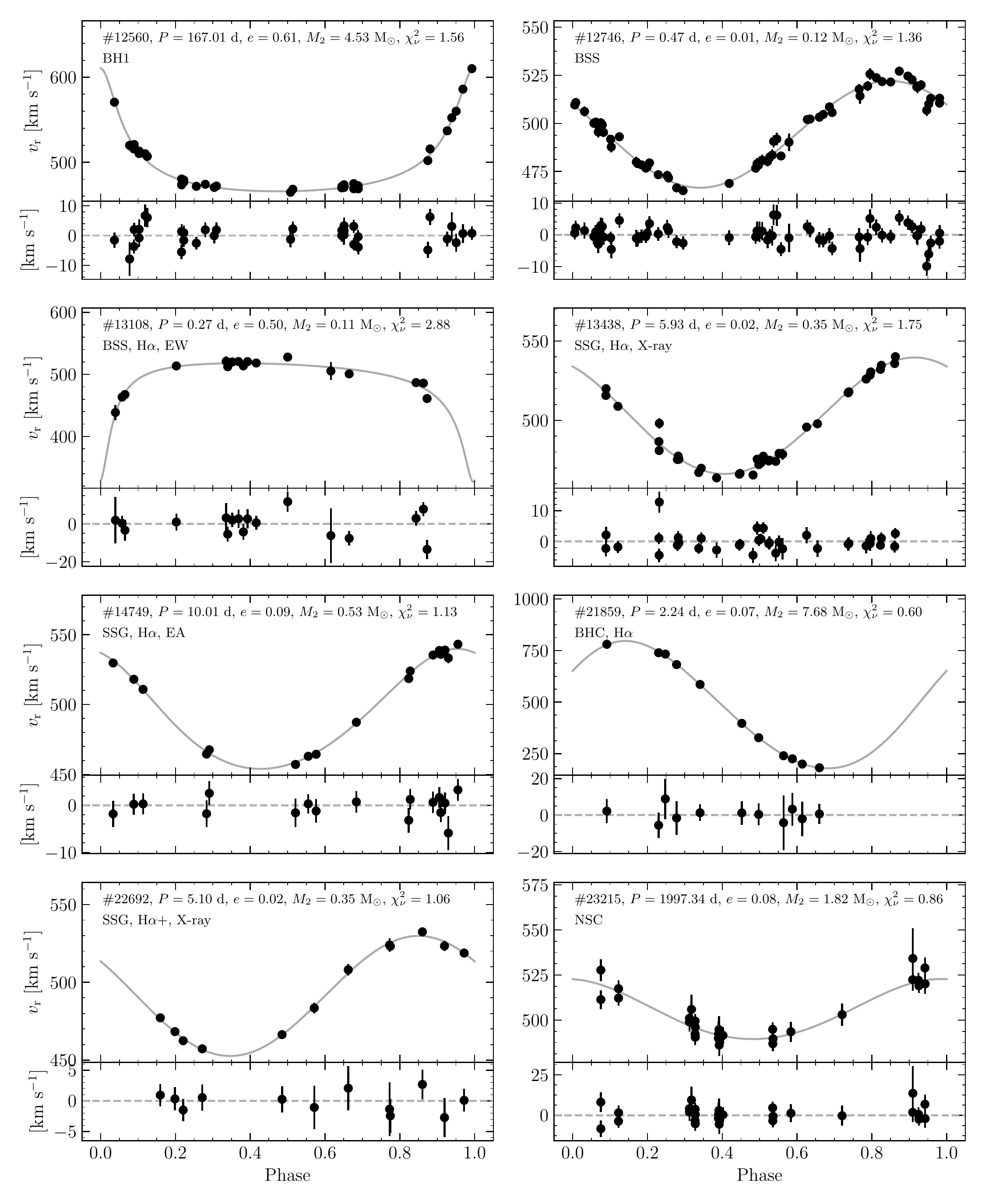}}
\caption{Selection of interesting results from \joker{} sorted by ACS Id. The upper panel of every plot shows the radial velocities of the star from our final sample phase folded with the period from the best-fitting model. (Sometimes the uncertainties appear smaller as the data points.) The best-fitting model is plotted with the continues line. The lower panel contains the residuals after subtracting this model from the data. We note the source number, period $P$, eccentricity $e$, invisible mass $M_2$ and reduced $\chi^2$ of the best fitting model in every plot. See \autoref{table:observations} for more information to individual stars.}
\label{fig:joker_interesting}
\end{figure*}

\begin{figure*}
\centering{\textbf{Random selection of Joker results}}
\resizebox{\hsize}{!}{\includegraphics{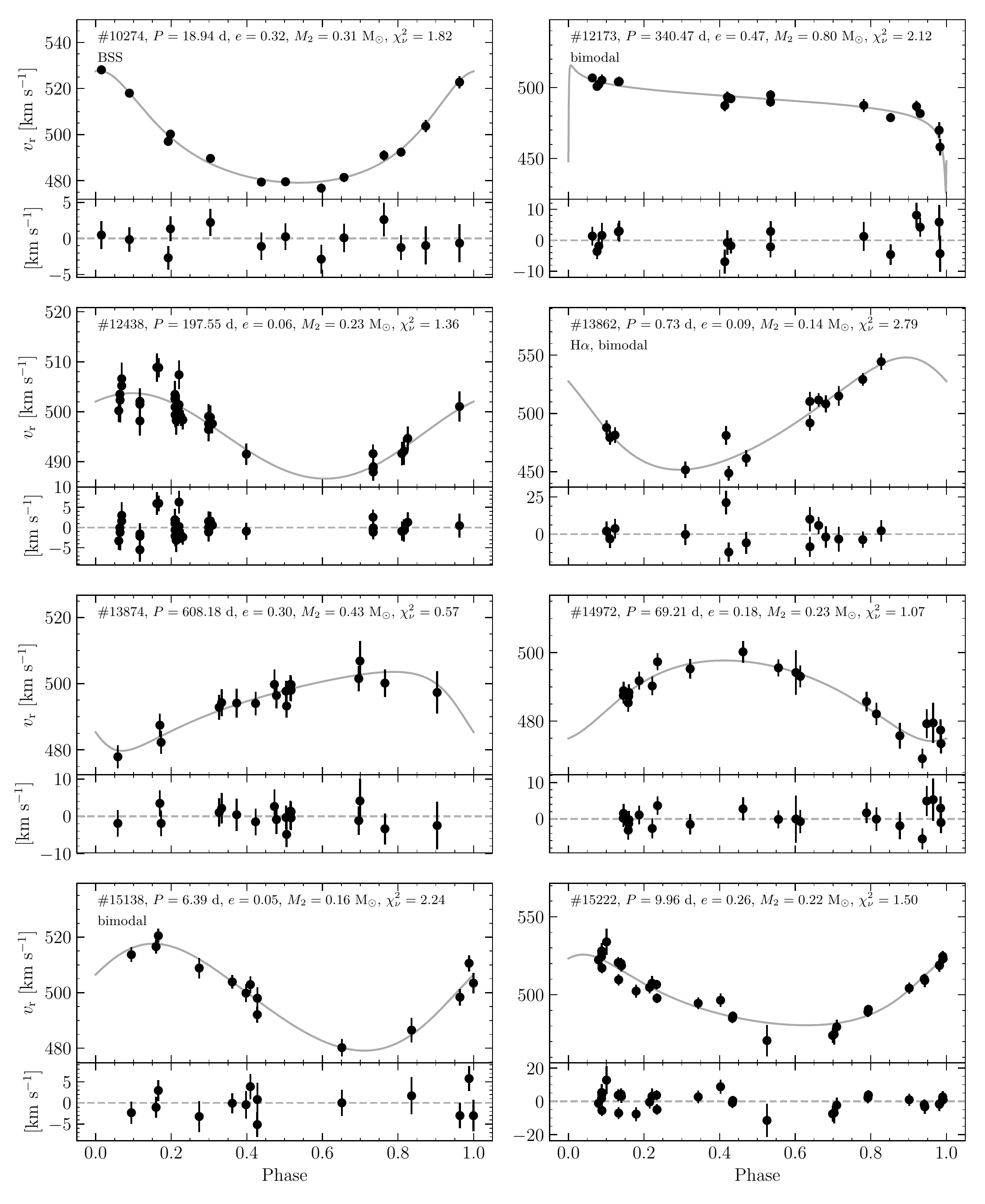}}
\caption{Same as \autoref{fig:joker_interesting} for a random selection of results from \joker{}.}
\label{fig:joker_random}
\end{figure*}

\begin{figure*}
\resizebox{\hsize}{!}{\includegraphics{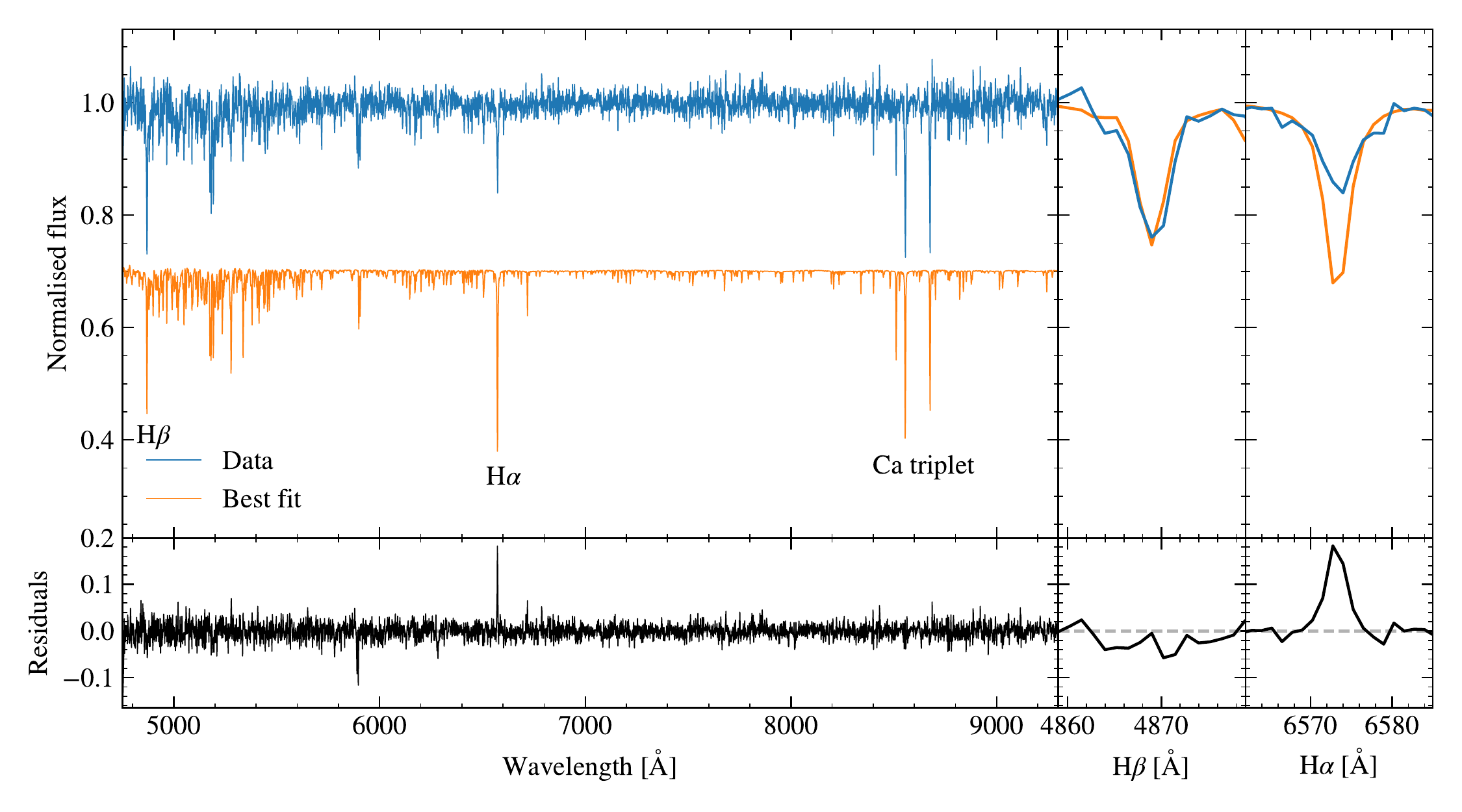}}
\caption{\changed{Spectrum of the \ssg{} with ACS Id \#13438. The top panel shows one normalised observed spectrum in blue and the best-fitting PHOENIX spectrum offset by 0.3 in orange. The bottom panel shows the residuals after subtracting the best fit from the data. The right panels show a zoom at the H$\beta$ and H$\alpha$ line with a filled-in observed H$\alpha$ line in contrast to the best fit (not offset).}}
\label{fig:ssg_filled}
\end{figure*}

\begin{figure*}
\resizebox{\hsize}{!}{\includegraphics{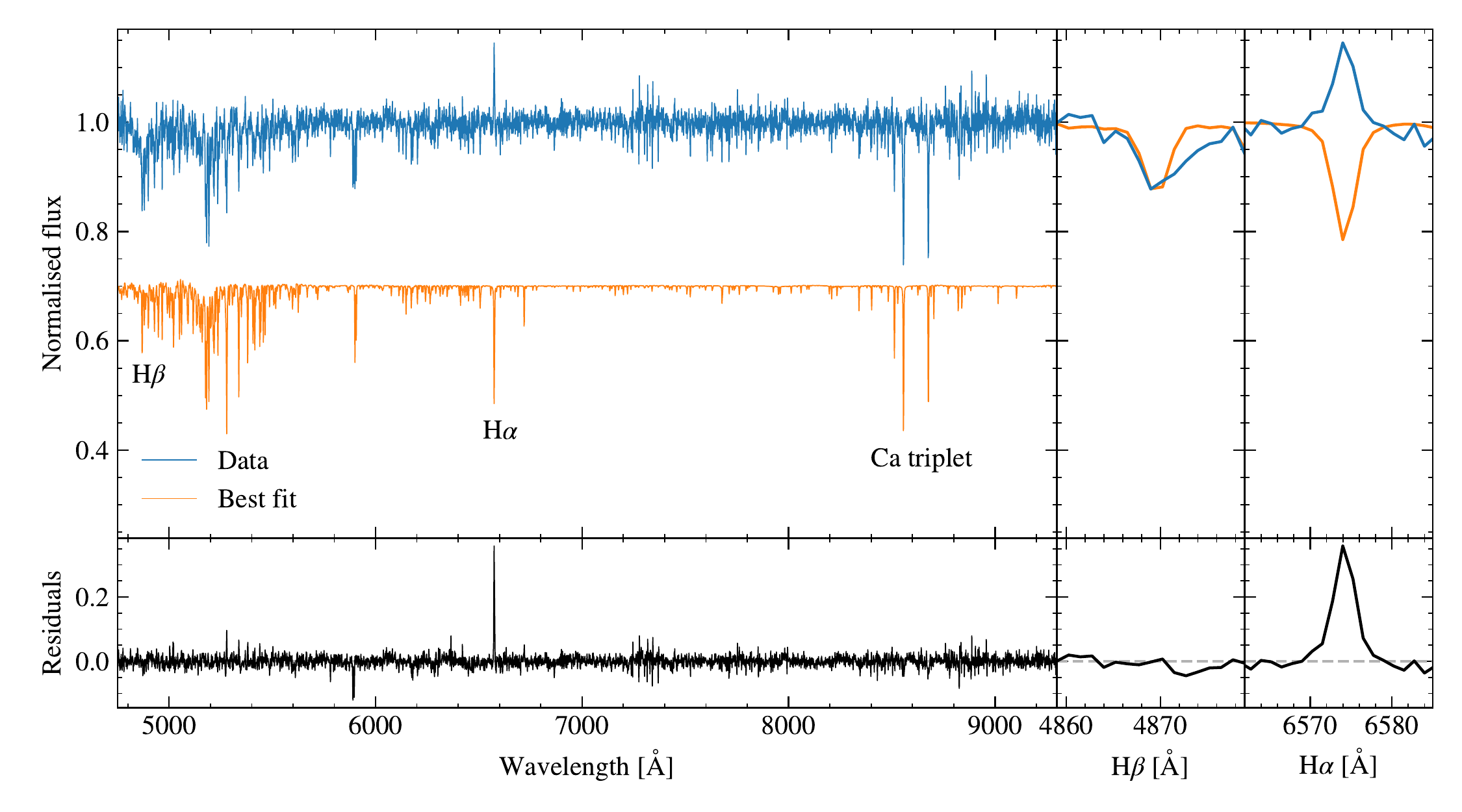}}
\caption{\changed{Same as Fig.~\ref{fig:ssg_filled} for the \ssg{} with ACS Id \#22692. Note that in contrast to star \#13438 (cf. \autoref{fig:ssg_filled}), this star shows an observed H$\alpha$ emission line.}}
\label{fig:ssg_emission}
\end{figure*}

\begin{figure*}
\resizebox{\hsize}{!}{\includegraphics{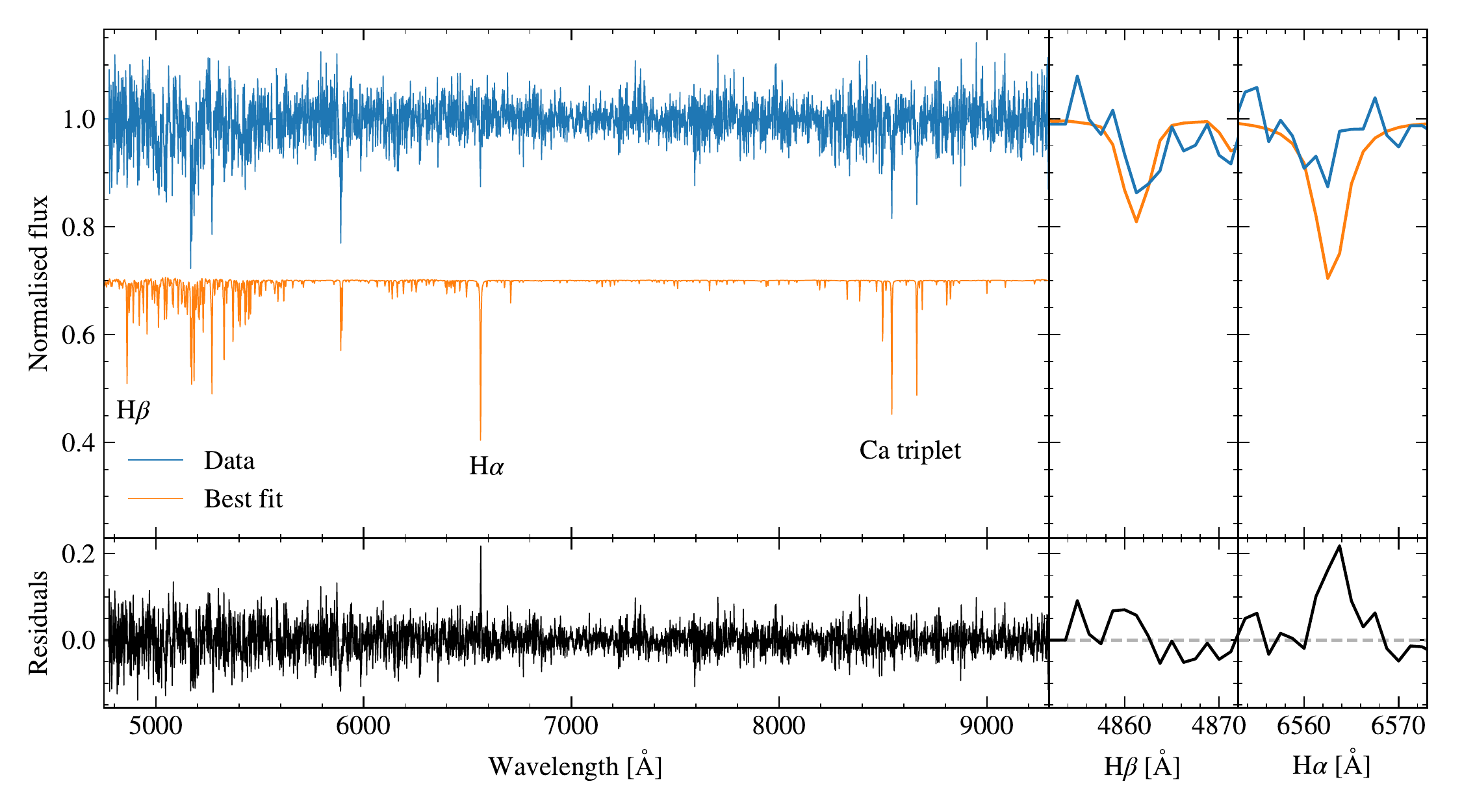}}
\caption{\changed{Same as Fig.~\ref{fig:ssg_filled} for the companion of the black hole candidate with ACS Id \#21859.}}
\label{fig:bh_candidate}
\end{figure*}